\documentclass[12pt,letterpaper]{article}
\usepackage{amsmath}
\usepackage{amsfonts}
\usepackage{algorithm}
\usepackage{verbatim}
\usepackage{geometry}  
\usepackage{graphicx}
\usepackage{caption}
\usepackage{mathtools}
\usepackage{multirow}
\usepackage{float}
\usepackage{amsthm}
\usepackage{amssymb}
\usepackage{epstopdf}
\usepackage{setspace}
\usepackage{fullpage}
\usepackage{enumerate}
\usepackage{authblk}
\usepackage{natbib}
\usepackage{hyperref}
\usepackage{bm}
\usepackage{color}
\usepackage{amsfonts} 
\usepackage{graphicx}
\usepackage{mathtools}
\usepackage{multirow}
\usepackage{bm}
\usepackage{hhline}
\usepackage{xcolor}
\usepackage{subcaption}
\usepackage{caption}

\makeatletter
\def\BState{\State\hskip-\ALG@thistlm}
\makeatother

\DeclareMathOperator*{\argmin}{\arg\!\min}
\DeclarePairedDelimiter{\ceil}{\lceil}{\rceil}

\newcommand{\bfxi}{{\bf x}_i}
\newcommand{\bfxki}{{\bf x}_{{\bf k}_i}}
\newcommand{\bfxkj}{{\bf x}_{{\bf k}_j}}

\newcommand{\bfz}{{\bf 0}}

\newcommand{\bfX}{{\bf X}}

\newcommand{\bfXk}{{\bf X_k}}

\newcommand{\bfk}{{\bf k}}
\newcommand{\bfy}{{\bf y}}

\newcommand{\con}{\,|\,}

\newcommand{\bfb}{{\boldsymbol \beta}}

\newcommand{\bfbk}{{\mathbf{\bm{\beta}_k}}}

\newcommand{\bfgk}{{\mathbf{\bm{\gamma}_k}}}

\newcommand{\bfdel}{{\boldsymbol \delta}}

\hypersetup{
	colorlinks,
	citecolor= blue
}

\graphicspath{{./Figures/}}

\begin{document}
\title{Bayesian Variable Selection For Survival Data Using Inverse Moment Priors}

%\author{%
%	Amir Nikooienejad\footremember{tamu}{Texas A\&M University}%
%	\and Walter Sobchak\footremember{mda}{MD Anderson Cancer Center}%
%	\and Valen E. Johnson\footrecall{tamu}% \footnote{Mexico?}%
%}

\author[1]{Amir Nikooienejad\thanks{Supported by NIH grant R01CA158113 prior to joining Eli Lilly and Company. \\Email: nikooienejad\_amir@lilly.com}}%
\author[2]{Wenyi Wang\thanks{Supported by 1R01CA174206, 1R01CA183793, 5R01CA158113 and P30CA016672.\\Email: wwang7@mdanderson.org}}%
\author[3]{Valen E. Johnson\thanks{Supported by NIH grant R01CA158113.\\Email: vjohnson@stat.tamu.edu}}%
\affil[1]{Department of Global Statistical Sciences, Eli Lilly and Company}
\affil[2]{Department of Bioinformatics and Computational Biology, University of Texas MD Anderson Cancer Center}
\affil[3]{Department of Statistics, Texas A\&M University}

\renewcommand\Authands{ and }
\renewcommand\Affilfont{\itshape\normalsize}
\date{}

\maketitle

\begin{abstract}
	Efficient variable selection in high-dimensional cancer genomic studies is critical for discovering genes associated with specific cancer types and for predicting response to treatment. Censored survival data is prevalent in such studies. In this article we introduce a Bayesian variable selection procedure that uses a mixture prior composed of a point mass at zero and an inverse moment prior in conjunction with the partial likelihood defined by the Cox proportional hazard model. The procedure is implemented in the R package BVSNLP, which supports parallel computing and uses a stochastic search method to explore the model space.  Bayesian model averaging is used for prediction. The proposed algorithm provides better performance than other variable selection procedures in simulation studies, and appears to provide more consistent variable selection when applied to actual genomic datasets.
\end{abstract}

\section{Introduction}
Recent developments in sequencing technology have made it easier to collect massive genomic datasets that can be used to study cancer and other diseases.  Given such data, there is great interest in linking genomic data to patient outcomes, and in many cases such outcomes are censored survival times. % The outcomes for most cancer studies are survival times for subjects, and the goal is to investigate the relation or any potential association between survival times and the covariates in the model; namely, genes in this context. %This is also known as survival regression problem that is, in concept, very close to regression analysis. 

Survival times for patients generally represent either the time to death or disease progression, the time to study termination or the time until the subject is lost to follow up. In the latter cases the subject's survival time is \emph{censored}. The relation between survival times and covariates is modeled through the hazard function, which is the limiting rate of death in the interval $(t,t+\Delta t)$ as $\Delta t$ becomes small, given patient covariates. More precisely, the hazard function $h$ for patient $i$ may be defined as

\begin{equation}\label{c4:hrate}
h(t\con\bfxi) = \lim\limits_{\Delta t \rightarrow 0}\frac{1}{\Delta t}P(t\leq T \leq t+\Delta t \con T\geq t, {\bf x}_i),
\end{equation}
where ${\bf x}_i$ is a $p$ vector of covariates thought to influence survival.  We denote by ${\bf X}$ the $n\times p$ design matrix obtained by stacking $n$ patient covariate vectors. 
Proportional hazard models take the form 
\begin{equation}\label{c4:hrate2}
h(t\con\bfxi) = h_0(t)\Phi(\bfxi)
\end{equation}
with an identifiability constraint of $\Phi(\bfz)=1$. In this formula, $h_0(t)$ denotes the baseline hazard function. The Cox proportional hazards model \citep{cox} is defined by taking $\Phi(\bfxi) = \exp\{\bfxi^T\bfb\}$, leading to
\begin{equation}\label{c4:link}
h(t\con\bfxi) = h_0(t)e^{\bfxi^T\bfb}.
\end{equation}
Here, $\bfb$ is a $p\times 1$ vector of coefficients. 

An important feature of the proportional hazards model is that it yields a partial likelihood function that is independent of the baseline hazard function, $h_0$. For complete survival analyses, however, the baseline hazard function is necessary for predicting survival times and can be estimated nonparametrically. Further details regarding the Cox proportional hazard model may be found in \citet{cox1984}, \citet{kalb} or \citet{cox}.

Gene expression datasets usually contain measurements on thousands of genes collected for only hundreds of subjects. Biologically, it seems plausible that only a relatively small number of these genes contribute significantly to survival. This implies that most of the elements in the vector $\bfb$ are small or close to zero. The challenge is to find covariates with nonzero coefficients or, equivalently, those genes that contribute the most in determining the survival outcome.

Many common penalized likelihood methods originally introduced for linear regression have been extended to survival data. These methods include LASSO \citep{lassosurvival}, in which an $L_1$ penalty is imposed on regression coefficients. \citet{adlassocox} utilized adaptive LASSO methodology for time to event data, while \citet{dantzigsurv} adopted the Dantzig selector for survival outcomes. The extension of nonconvex penalized likelihood approaches, in particular SCAD, to the Cox proportional hazard model is discussed in \citet{scadsurvival}. The Iterative Sure Independence Screening (ISIS) approach introduced by \citet{isis} is also extended for ultrahigh dimensional survival data in \citet{fansurvisis}, where it is used on Cox proportional hazard models and the SCAD penalty is employed for variable selection.

Some Bayesian approaches have also been introduced. \citet{bayes1} proposed a method based on approximating the posterior distribution of the parameters in the proportional hazard model by defining a Gaussian prior on regression coefficients. A loss function was then imposed to select a parsimonious model. A semiparametric Bayesian approach was utilized by \citet{bayes2}, who employed a discrete gamma process for the baseline hazard function and a multivariate Gaussian prior for the coefficient vector. \citet{bayes3} considered Accelerated Failure Time (AFT) models along with data augmentation to impute failure times. A mixture prior proposed by \citet{georgeapproaches} was used to impose sparsity. In more recent work, \citet{leo} proposed the use of a $g$-prior  model for the coefficient vector and employed test-based Bayes factors \citep{valtbf} to the Cox proportional hazard models. However, this method is intended for use only when the number of covariates is less than the number of observations, that is, when $p < n$.

%Due to the huge computational load of Bayesian data analysis imposed by Monte Carlo Markov Chain (MCMC) procedure, in particular for high dimensional survival data, the frequentist approaches outnumber their Bayesian counterparts in real genomic applications. Consequently, developing a fairly fast Bayesian variable selection method for high dimensional datasets that can outperform dominant frequentist algorithms seems compelling.

To our knowledge, all previous Bayesian procedures for variable selection in survival data have used local priors on model coefficients. In Bayesian hypothesis tests, local priors put a positive probability on the null value of the parameter, in this case zero, whereas nonlocal priors put zero probability on the null value. \cite{nonlocalprior} can be consulted for more discussion on properties of local and nonlocal priors in the context of Bayesian testing. In this article we propose a Bayesian method based on a mixture prior comprised of a point mass at zero and a nonlocal prior on the regression coefficients. To handle the computational burden of implementing the resulting procedure, we employ a stochastic search method, S5 \citep{s5}, which we implement in an R package BVSNLP.  We also discuss a general procedure for setting the tuning parameter of the nonlocal prior. 

This article is structured as follows. In Section \ref{c4:model} we introduce notation and discuss the modeling of the problem in a Bayesian framework. Section \ref{c4:method} discusses the proposed method, with details of parameter selection, model search and assessment of the accuracy of the proposed variable selection procedure. Sections \ref{c4:resultsimul} and \ref{c4:resultreal} provide simulation and real data analyses with various predictive performance measures to demonstrate how the proposed method compares to several other competing methods. Section \ref{c4:discuss} concludes with discussion.
%================================================================================================
%================================================================================================

\section{Problem Modeling}\label{c4:model}
\subsection{Preliminaries}
Let $T_i$ denote the survival time and $C_i$ denote the censoring time for individual $i$. Each element in the observed vector of survival times, $\bfy$, is defined as $y_i=\min\{T_i,C_i\}$. The status for each individual is defined as $\delta_i=I(T_i\leq C_i)$. The status vector is represented by $\bfdel = (\delta_1,\delta_2,\ldots,\delta_n)^T$. We assume that the censoring mechanism is ``at random,'' meaning that $C_i$ and $T_i$ are conditionally independent given $\bfxi$, where $\bfxi \in \mathbb{R}^p$ are the covariates for individual $i$ and comprise the $i^{th}$ row of $\bfX$. The observed data is of the form $\big\{(y_i,\delta_i,\bfxi); \; i=1,2,\ldots,n\big\}$.

Model $\bfk$ is defined as $\mathbf{k}=\{k_1,\ldots ,k_j\}$, where $(1\leq k_1 < \cdots < k_j \leq p)$, and it is assumed that $\beta_{k_1}\neq 0, \ldots , \beta_{k_j}\neq 0$ and all other elements of $\bm{\beta}$ are 0. The design matrix corresponding to model $\bfk$ is denoted by $\bfXk$ and the regression vector by $\bfbk = (\beta_{k_1},\beta_{k_2},\ldots,\beta_{k_j})^T$.

Let $\mathcal{R}(t)=\{i:y_i\leq t\}$ represent the \emph{risk set} at time $t$, the set of all individuals who are still present in the study at time $t$ and are neither dead nor censored. We assume throughout this article that the failure times are distinct. In other words, only one individual fails at a specific failure time. With this assumption and letting $\xi_{\bfk_i}=\exp\{\bfxki^T\bfbk\}$, the partial likelihood \citep{cox} for $\bfbk$ in model $\bfk$ can be written as

\begin{equation}\label{c4:partial}
L_p(\bfbk)=\prod\limits_{i=1}^n\Bigg[\frac{\xi_{\bfk_i}}{\sum\limits_{j\in R(y_i)}\xi_{\bfk_j}}\Bigg]^{\delta_i}.
\end{equation}

Our method uses this partial likelihood as the sampling distribution in our Bayesian model selection procedure. We acknowledge that there is some information loss in (\ref{c4:partial}) with respect to $\bfbk$. For instance, Basu \citep{gosh} argues that partial likelihoods cannot usually be interpreted as sampling distributions. On the other hand, \citet{berger} encourage the use of partial likelihoods when the nuisance parameters are marginalized out. %We chose to test this idea and use the partial likelihood in (\ref{c4:partial}) as if it was the sampling distribution for observed survival times.

Sorting the observed unique survival times in ascending order and, consequently, reordering the status vector $\bfdel$ as well as the design matrix $\bfX$ with respect to the ordered $\bfy$, the sampling distribution of $\bfy$ for model $\bfk$ can be written as

\begin{equation}\label{c4:sampling}
\pi(\bfy\con\bfbk)=\prod\limits_{i=1}^n\Bigg[\frac{e^{\bfxki^T\bfbk}}{\sum\limits_{j=i}^n e^{\bfxkj^T\bfbk}}\Bigg]^{\delta_i}.
\end{equation}  

A Bayesian hierarchical model can be defined in which $\pi(\bfy\con\bfbk)$ in (\ref{c4:sampling}) represents the sampling distribution, $\pi_\bfk(\bfbk)$ is the prior of model coefficients $\bfbk$ and $P(\bfk)$ is the prior for model $\bfk$. Using Bayes rule, the posterior probability for model ${\bf j}$ is written as

\begin{equation}\label{c4:posterior}
P(\mathbf{j}\con \bfy)=\frac{P(\mathbf{j})m_{\mathbf{j}}(\bfy)}{\sum_{k\in \mathcal{J}} P(\mathbf{k})m_{\mathbf{k}}(\bfy)},
\end{equation}
where $\mathcal{J}$ is the set of all possible models and the marginal probability of the data under model $\bfk$ is defined by
\begin{equation}\label{c4:marginal}
m_{\mathbf{k}}(\bfy)=\int\pi(\bfy \con \bfbk)\pi_\bfk(\bfbk)d\bfbk.
\end{equation}

The prior density for $\bfbk$ and the prior on the model space impact the overall performance of the selection procedure and the amount of sparsity imposed on candidate models. Note that the sampling distribution in (\ref{c4:sampling}) is continuous in $\bfbk$, and in Section~2.3 we define an inverse moment prior \citep{nonlocalprior} on each of the coefficients in model $\bfk$.

\subsection{Prior on Model Space}\label{c4:pms}
Let $\bfgk=\{\gamma_1,\cdots,\gamma_p\}$ denote a binary vector indicating which covariates are included in model $\bfk$. Suppose the size of model $\bfk$ is $k$. That is, there are $k$ nonzero indices in $\bfgk$. The nonzero indices of $\bfgk$ represent the indices of the nonzero elements in the coefficient vector, $\bfb$, which a priori are modeled as independent Bernoulli random variables with success probability $P(\gamma_i=1)=\theta$ for every $1 \leq i \leq p$. As discussed in \citet{scott2010}, no fixed value for $\theta$ adjusts for multiplicity. As a result, it is necessary to define a prior on $\theta$, say $\pi(\theta)$. The resulting marginal probability for model $\bfk$ in a fully Bayesian approach may then be written as

\begin{equation}\label{c2:priormod}
p(\bfk) \propto \int \theta^k(1-\theta)^{p-k}\pi(\theta) d\theta.
\end{equation}

A common choice for $\pi(\theta)$ is the beta distribution, $\theta\sim\text{Beta}(a,b)$, where in the special case of $a=b=1$, $\pi(\theta)$ is a uniform distribution. The marginal probability for model $\bfk$ derived from (\ref{c2:priormod}) is then equal to
\begin{equation}\label{c2:priormod2}
p(\bfk) = \frac{B(a+k,b+p-k)}{B(a,b)},
\end{equation}
where $B(\cdot)$ is the Beta function. {\em A priori,} the model size, $k$, thus follows a Beta-binomial distribution. By choosing $b=p-a$, the mean and variance of the selected model size $k$ is

\begin{equation}\label{c2:modsizestat}
\mathbb{E}(k)=a, \quad \text{Var}(k)=\frac{2p^2a(p-a)}{p^3}%\approx 2(a-\frac{a^2}{p})
\approx 2a.
\end{equation}

The approximation in the variance formula follow from a large $p$ and a fairly small $a$ under the sparsity assumption on the true model size. To incorporate the belief that the optimal predictive models are sparse, we recommend setting $a=1$ and $b=p-a$.  The resulting prior assigns comparatively small prior probabilities to models that contain many covariates. %However, in some situations, depending on the penalization provided by the prior on coefficients and the nature of the problem, the uniform-binomial might be the preferred choice. For instance, if the true model is assumed to be really sparse, beta-binomial prior is preferred but if not enough information is available regarding the true model size, our recommendation is to use uniform-binomial prior where less penalty is imposed on the model size.

\subsection{Product Inverse MOMent (piMOM) Prior}\label{c4:pimommethod}
\begin{figure}[!t]
	\centering
	\includegraphics[width=80mm]{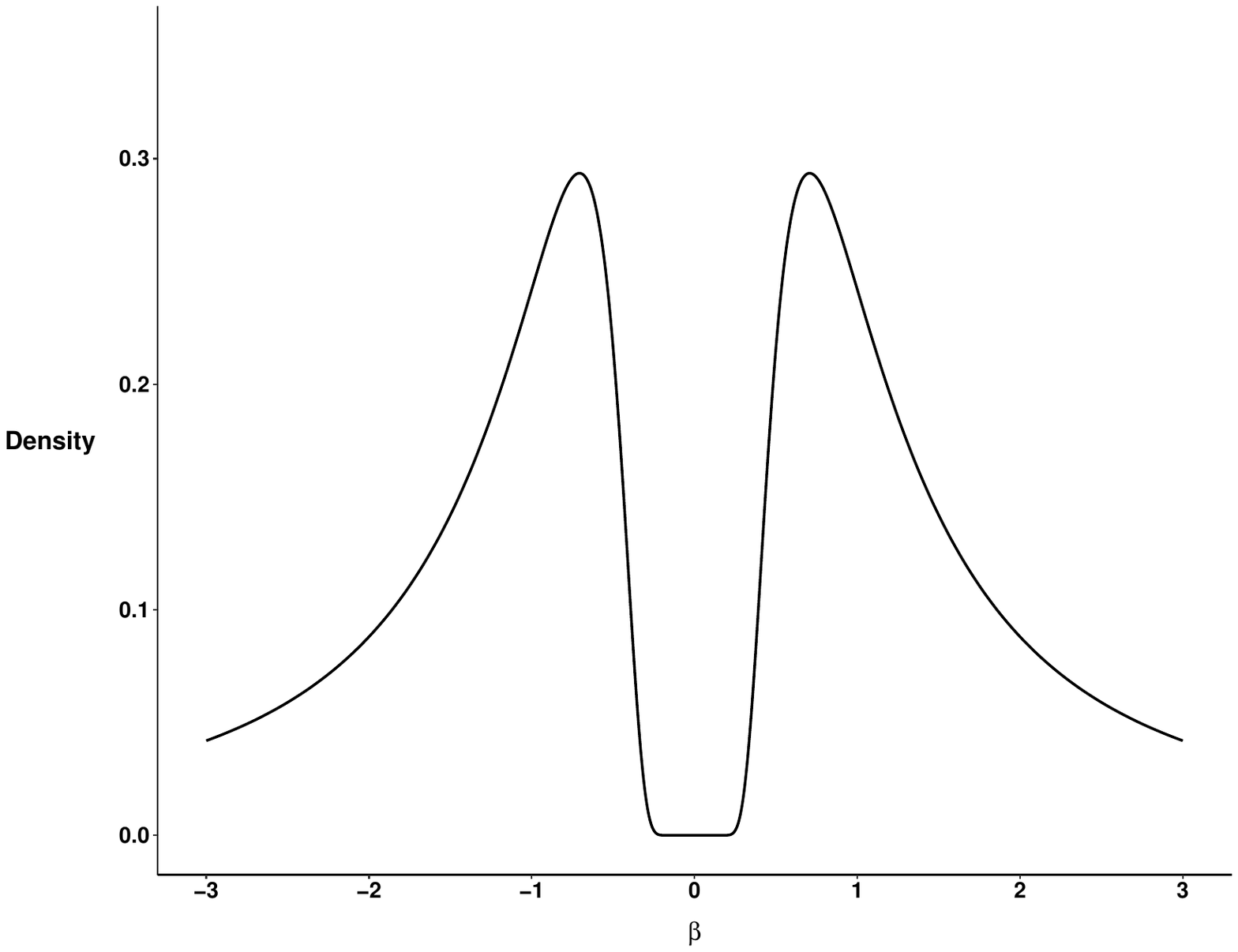}
	\caption{iMOM prior with $r=1$ and $\tau=0.5$.}\label{pimom_only}
\end{figure}

We impose nonlocal prior densities on the nonzero coefficients, $\bfbk$. Specifically, we assume the prior densities on the nonzero coefficients in model $\bfk$ take the form of a product of independent  iMOM priors, or piMOM densities \citep{johnsonrossel}, expressible as
\begin{equation}\label{c4:pimom}
\pi(\bfbk|\tau,r)=\frac{\tau^{rk/2}}{\Gamma(r/2)^k}\prod_{i=1}^{k}|\beta_i|^{-(r+1)}\exp\Big(-\frac{\tau}{\beta_i^2}\Big) , \qquad r,\tau>0.
\end{equation}
The hyperparameter $\tau$ represents a scale parameter that determines the dispersion of the prior around $\bfz$, while $r$ determines the tail behavior of the density. These priors have two symmetric modes with Cauchy-like tails when $r=1$, and assign negligible probability to a region around zero. In comparison to local priors, this characteristic of nonlocal priors potentially leads to smaller false positive rates in selection procedures by discouraging the selection of variables with small coefficients. On the other hand, piMOM priors possess Cauchy-like tails which introduce comparatively small penalties on large coefficients. Unlike many penalized likelihood methods, large values of regression coefficients are thus not heavily penalized by these priors. As a result, they do not necessarily impose significant penalties on nonsparse models provided that the estimated coefficients in those models are not small.  For these reasons piMOM priors work well as a default choice of priors on nonnegligible coefficients in variable selection problems. An example of an iMOM prior is depicted in Figure \ref{pimom_only} for $r=1$ and $\tau=0.5$.

Another nonlocal prior that might be considered as a potential candidate for the prior densities on the nonzero coefficients in model $\bfk$ is the product of independent MOM priors, or the pMOM densities \citep{johnsonrossel}. A detailed discussion on the pMOM priors and their properties is provided in Section 3 of the \ref{suppA} \citep{supplement}. A simulation analysis to compare the performance of piMOM and pMOM in the selection process is also provided. For the reasons discussed there, piMOM-based procedures are more effective for variable selection in $p\gg n$ settings and, therefore, is our choice for the analyses of the simulation and real data in this article.

\section{Methods}\label{c4:method}
\subsection{Selection of Hyperparameters}\label{hypersel}
We use the procedure described in \citet{imomlogit} to select hyperparameter values for the piMOM prior. In that method, the null distribution of the maximum likelihood estimator for $\bfbk$ (i.e., all components of $\bfbk$ are 0), obtained from randomly selected design matrices $\bfXk$, is compared to the prior density on $\bfbk$ for various values of $(r,\tau)$. Fixing $r=1$ to achieve Cauchy-like tails, a value of $\tau$ is chosen so that the overlap between the two densities is less than a specified threshold, $1/\sqrt{p}$, and is denoted by $\tau_1$. It can be shown that the maximum of the iMOM prior occurs at $\pm\sqrt{\tau}$. We also allow users to input a prior parameter $\alpha$ that controls where the modes in the prior occur. This can be useful in constraining the prior density when covariates are highly correlated (resulting in an over-dispersed prior when the sampling distribution of the null MLE under the null model becomes overly broad). We then set the value of $\tau$ according to

\begin{equation}\label{hypertau}
\tau=\min(\tau_1,\alpha^2).
\end{equation}
%Increasing $\tau$ in the iMOM prior widens the zero probability region around the null value. Therefore, we use the minimum between $\tau_1$ and $\alpha$ in (\ref{hypertau}) to capture the desired effect size as well as accounting for $1/\sqrt{p}$ overlap rule.

To implement the procedure for computing $\tau_1$ for survival models, we generate response vectors under the null model using the procedure described by \citet{bender}.  Survival times are sampled from a standard exponential model.   %when all components of $\bfb$ are zero. As a result, for each individual, the sampled survival time under the null is computed as

%\begin{equation}\label{c4:samplest}
%t^s_i=-\frac{\log{u_i}}{\lambda_1\exp\{\bfX\bfb\}} = -\frac{\log{u_i}}{\lambda_1};\quad \mbox{ where } u_i\sim U(0,1).
%\end{equation}
%In this formulation, $\lambda_1$ is the baseline hazard function, $h_0(t)$, which we assume to be $1$ in our analysis, and $U(0,1)$ is the uniform distribution between $0$ and $1$. I define the event rate to be the proportion of subjects that have $\delta_i=1$. This can be estimated from observed data by taking the average of the event status vector, $\bfdel$. Defining censoring rate as one minus the even rate, The estimated censoring rate can be obtained by
%\begin{equation}\label{c4:defcensrate}
%\hat{c} = 1-\hat{e},
%\end{equation}
%where $\hat{e}$ and $\hat{c}$ are the estimated event and censoring rate, respectively.

Let $\bm{t}^s$ and $\bm{c}^s$ be the vector of sampled survival times and censoring times, respectively.
%
%The censoring times, are obtained independently by sampling from an exponential distribution with rate $\lambda_2$. The rate $\lambda_2$ is computed from the assumed rate for exponential distribution in sampling survival times in (\ref{c4:samplest}), $\lambda_1$, and the estimated censoring rate in observations, $\hat{c}$. More precisely, $\lambda_2$ is set so the censoring rate is equal to the observed censoring rate, $\hat{c}$. The details of this calculation are described by the following equation:
%
%\begin{equation}\label{c4:estcrate}
%\hat{c} = \mathbb{E}\big[I(t_i^s > c_i^s)\big] = p(t_i^s > c_i^s) = \int_{0}^{\infty}\int_{c}^{\infty}\lambda_1\lambda_2e^{-\lambda_1t}e^{-\lambda_2c}dtdc = \frac{\lambda_2}{\lambda_1+\lambda_2}  .
%\end{equation}
%
%Thus, letting $\lambda_1=1$, the rate $\lambda_2$ is computed as
%
%\begin{equation}\label{c4:rate4c}
%\lambda_2 = \frac{\hat{c}}{1-\hat{c}},
%\end{equation}
%which leads us to obtaining censoring times vector, $\bm{c}^s$.
The sampled survival time and status for each observation is then computed as
\begin{equation}
y_i^s = \min\{t_i^s,c_i^s\} \quad \quad \text{ and } \quad \quad  \delta_i^s = I(t_i^s \leq c_i^s),
\end{equation}
which comprise $\bfy^s$ and $\bfdel^s$ under the null model. Using the pair $(\bfy^s, \bfdel^s)$, the MLE from a Cox model is computed. It should be noted that the asymptotic distribution of the MLE for the Cox model under the null hypothesis is $\hat{\bfb}\sim \mathcal{N}\big(\bfz,I(\hat{\bfb})\big)$, where $I(\bfb)$ is the information matrix of the partial likelihood function. Thus, it is appropriate to approximate the pooled estimated coefficients in that algorithm with a normal density function. When the sample size gets large, the variance of the MLE decreases and causes the overlap to become small and, consequently, small values of $\tau$ are selected.

In general, we find that $r=1$ and $\tau=0.25$ are good default values if one chooses not to run the hyperparameter selection algorithm. When $r=1$, the peaks of the iMOM prior occur at $-\sqrt{\tau}$ and $\sqrt{\tau}$. By equating $\sqrt{\tau}$ to the absolute value of the expected effect size for a given application, insight can be gained on what value of $\tau$ is appropriate. Further details regarding this algorithm can be found in \citet{imomlogit}.

\subsection{Computing Posterior Probability of Models}
Computing the posterior probability for each model requires the marginal probability of observed survival times under each model, as shown in (\ref{c4:posterior}) and (\ref{c4:marginal}). The marginal probability is approximated using the Laplace approximation, where the regression coefficients in $\bfbk$ are integrated out. This leads to

\begin{equation}\label{marginal}
m_{\bfk}(\bfy_n) = \pi(\bfy_n\con \hat{\bfb}_\bfk)\pi(\hat{\bfb}_\bfk) (2\pi)^{k/2}|G_{\hat{\bfb}_\bfk}|^{-1/2}.
\end{equation}

Here, $\bm{\hat{\bfb}_\bfk}$ is the maximum a posteriori (MAP) estimate of $\bfbk$, $G_{\hat{\bfb}_\bfk}$ is the Hessian of the negative of the log posterior function,

\begin{equation}\label{logpost}
g(\bfbk)=-\log(\pi(\bfy\con\bfbk))-\log(\pi(\bfbk)), 
\end{equation}
computed at $\bm{\hat{\bfb}_\bfk}$ and $k$ is the size of model $\bfk$. Finding the MAP of $\bfbk$ is equivalent to finding the minimum of $g(\bfbk)$.

The details of computing the gradient and Hessian matrix of $g(\bfbk)$ are discussed in Section 1 of the \ref{suppA} \citep{supplement}. The gradient and Hessian matrix, described by equations (3) to (7) there, are used to find the MAP and to compute the Laplace approximation of the marginal probability of $\bfy$. We use the limited memory version of the Broyden--Fletcher--Goldfarb--Shanno optimization algorithm (L-BFGS) \citep{lbfgsref} to find the MAP. The initial value for the algorithm is $\hat{\bfb}_\bfk$, the MLE for the Cox proportional hazard model. 

Having all the components of formula (\ref{c4:posterior}), it is possible to define a MCMC framework to sample from the posterior distribution on the model space. A birth-death scheme, similar to that used in \citet{imomlogit}, could be used for this purpose. However, for computational reasons we use another stochastic algorithm to search the model space; this algorithm is described in the next section. %As a result, an alternative technique, discussed in the following, is utilized in order to bypass burdensome MCMC procedure and make the algorithm computationally feasible.

The highest posterior probability model (HPPM) is defined as the model having the highest posterior probability among all visited models. In practice, many models may be assigned probabilities that are close to the probability achieved by the HPPM. For this reason and for predictive purposes, it is useful to obtain the Median Probability Model (MPM) \citep{barbieri} which is the model containing covariates that have posterior inclusion probabilities of at least $0.5$. According to \citet{barbieri}, the posterior inclusion probability for covariate $i$ is defined as

\begin{equation}\label{c4:postincprob}
p_i = \sum_{\bfk:\; \bfgk_i = 1}P(M_\bfk|\bfy).
\end{equation}
That is, the sum of posterior probabilities of all models that have covariate $i$ as one of their variables. In this expression, $\bfgk_i$ is a binary value determining the inclusion of the $i^{th}$ covariate in model $\bfk$.

\subsubsection{Stochastic Search Algorithm}\label{c4:as5}
To increase the efficiency of exploring the model space, we use the S5 algorithm. S5 was proposed by \citet{s5} for variable selection in linear regression problems, and we adapt it here for survival models. It is a stochastic search method that screens covariates at each step. The algorithm is scalable and its computational complexity is only linearly dependent on $p$ \citep{s5}.

Screening is the essential part of the S5 algorithm. In linear regression, screening is based on the correlation between excluded covariates and the residuals of the regression using the current model \citep{isis}. The concept of screening covariates for survival response data is proposed in \citet{fansurvisis} and is defined based on the marginal utility for each covariate.

To illustrate the screening technique, suppose that the current model is $\bfk$. Let $\bfk^c$ denote the complement of set $\bfk$ containing columns of the design matrix that are not in the current model, $\bfk$. The conditional utility of covariate $m\in\bfk^c$ represents the amount of information covariate $m$ contributes to the survival outcome, given model $\bfk$, and is defined as
\begin{equation}\label{c4:condutil1}
\begin{split}
u_{m\con\bfk} = \max\limits_{\substack{\beta_m\\ m\in\bfk^c}}\bfdel^T\Big[&(\beta_m \bfX_{(m)}+\bfXk\bfbk)\\
& -\log\Big\{\sum\limits_{j=i}^n
\exp(\beta_m x_{jm}+\bfxkj\bfbk)\Big\}\Big].
\end{split}
\end{equation}
By comparing $u_{m\con\bfk}$ to the Cox log-likelihood equation (Formula (1) in the \ref{suppA} \citep{supplement}), it follows heuristically that the conditional utility is the maximum likelihood for covariate $m$ after accounting for the information provided by model $\bfk$. Finding $u_{m\con\bfk}$ is a univariate optimization procedure that can be computed rapidly.  

With this background the S5 algorithm for survival data works as follows. At each step the $d$ covariates with highest conditional utility are candidates to be added to the current model $\bfk$ and comprise the addition set, $\Gamma^+$. The deletion set, $\Gamma^-$ contains the current model, except that one variable is removed. From the current model, $\bfk$, we consider moves to each of its neighbors in $\Gamma^+$ and $\Gamma^-$ with a probability proportional to the marginal probabilities of these neighboring models.

To avoid local maxima, the model probabilities used in S5 are raised to the power of $1/t_l$, where $t_l$ is the $l^{th}$ temperature in an annealing schedule in which ``temperatures'' decrease. To increase the number of visited models, a specified number of iterations are performed at each temperature. At the end of the procedure, the model with the highest posterior probability of visited models is identified as the HPPM.

In our version of the S5 algorithm, we used $10$ equally spaced temperatures varying from $3$ to $1$ and $30$ iterations within each temperature. Section 4 of the \ref{suppA} \citep{supplement} provides some discussion on how these values are chosen for this application. To increase the number of visited models, we parallelized the S5 procedure so that it could be distributed to multiple CPUs. Each CPU executes the S5 algorithm independently with a different starting model. All visited models are pooled together at the end, and the HPPM and MPM are determined. Using posterior probabilities of the visited models, the posterior inclusion probability for each covariate can be computed using (\ref{c4:postincprob}). In our simulations we used $120$ CPUs to explore the model space for design matrices with $O(10^4)$ covariates.

%The proposed algorithm of Bayesian variable selection for survival data is implemented in the same R package with the methods for binary data discussed in Chapter \ref{Chapter:softpack}.

\subsection{Predictive Accuracy Assessment}\label{predictive}
In addition to looking at the selected genes and their pathways to determine their biological relevance in analyzing the real datasets, we used the time dependent AUC, obtained from time dependent ROC curves as introduced by \citet{roct}, for survival times to summarize and compare the predictive performance of the various algorithms. This measure has a relatively straightforward interpretation and, unlike other summary measures such as the c-index \citep{harrell}, can be computed without requiring specific conditions or additional assumptions to hold \citep{blanche}. However, predictive performance measures including the c-index, Integrated Brier Score (IBS)\citep{gerds} and prediction error curves, are investigated and reported in Sections \ref{c4:resultsimul} and \ref{c4:resultreal} for both simulation and real data sets.

There are different methods to estimate time dependent sensitivity and specificity. In our algorithm we adapted a method proposed by \citet{survuno}, henceforth called Uno's method. In that method, after splitting data into training and test sets, sensitivity is estimated by

\begin{equation}\label{se}
\widehat{\text{SE}}_{\bfk}(t,c) = \frac{\sum_{i=1}^{n}\delta_iI(\bfxki^T\hat{\bfb}_{\bfk} > c, T_i \leq t)/\hat{G}(T_i)}{\sum_{i=1}^{n}\delta_iI(T_i\leq t)/\hat{G}(T_i)},
\end{equation}
and specificity is estimated by
\begin{equation}\label{sp}
\widehat{\text{SP}}_{\bfk}(t,c) = \frac{\sum_{i=1}^{n}I(\bfxki^T\hat{\bfb}_{\bfk} \leq c, T_i > t)}{\sum_{i=1}^{n}I(T_i > t)}.	
\end{equation}

These values are estimated for the test set. Therefore, in the equations above, $n$ is the number of observations in the test set, $\delta_i$ is the status of observation $i$ and $T_i$ is the observed time for that observation in the test set. The variable $c$ is the discrimination threshold that is varied to obtain the ROC curve. The function $\hat{G}$ is the Kaplan--Meier estimate of the survival function obtained from the training set. For each observation $i$ in the test set with observed time $T_i$, $\hat{G}(T_i)$ is computed by a basic interpolation procedure. That is, 
\begin{equation}\label{gintpl}
\hat{G}(T_i) = \hat{G}(T_{tr}^*),\; \text{ where } T_{tr}^* = \argmin_{T\in T_{tr}} |T-T_i|.
\end{equation}
Here, $T_{tr}$ is the set of all observed survival times in the training set. In (\ref{se}) and (\ref{sp}), $\hat{\bfb}$ represents the estimated coefficient under a specific model. %This is where different models can play role in determining sensitivity and specificity.

\subsubsection{Bayesian Model Averaging (BMA)}
%In analyzing real datasets, to incorporate the uncertainty in the selected models, and the fact that only limited number of models are being visited in the selection procedure, it is compelling to use Bayesian model averaging for predictive purposes. In particular, we account for different models in computing time dependent sensitivity and specificity, and not just HPPM. Following (\ref{se}) and (\ref{sp}), let the final sensitivity and specificity using BMA is obtained by,
BMA can be used to improve the predictive accuracy by accounting for the uncertainty in selected models. From (\ref{se}) and (\ref{sp}) the final sensitivity and specificity using BMA may be defined as

\begin{equation}\label{se_bma}
\widehat{\text{SE}}_{\text{BMA}}(t,c) = \sum\limits_{j=1}^\mathcal{N} \widehat{\text{SE}}_{\bfk^j}(t,c) P(\mathcal{M}_{\bfk^j}\con \bfy_n),
\end{equation}
and
\begin{equation}\label{sp_bma}
\widehat{\text{SP}}_{\text{BMA}}(t,c) = \sum\limits_{j=1}^\mathcal{N} \widehat{\text{SP}}_{\bfk^j}(t,c) P(\mathcal{M}_{\bfk^j}\con \bfy_n),
\end{equation}
where, $P(\mathcal{M}_{\bfk^i}\con \bfy_n)$ is the posterior probability of model $\mathcal{M}_{\bfk^i}$ The value of $\mathcal{N}$ depends on what type of BMA is used. We use Occam's window, which means only models that have posterior probability of at least $w \times p(\mathcal{M}_{\text{HPPM}}\con \bfy_n)$ are used in model averaging. We set $w = 0.01$ for our applications.

In the proposed method individual survival curves are estimated using the highest posterior probability model. Section 2 of the \ref{suppA} \citep{supplement} provides the details of this procedure. Similar approaches were also adopted by \citet{leo} in estimating the survival curve for each individual in a study.

%================================================================================================
%================================================================================================

\section{Simulation Results} \label{c4:resultsimul}

\sloppy To investigate the performance of the proposed model selection procedure, we applied our method to simulated datasets. We followed the guidance of \citet{morris} as a basis for our simulation protocol. In particular, the simulation design was based on the ADEMP structure (Aims, Data generating mechanism, Estimands, Methods, and Performance measures) discussed in that article. We refer to each of those elements as we explain different parts of the simulation design below.

Regarding \emph{``Methods,''} we compared the performance of our algorithm to ISIS-SCAD \citep{fansurvisis} and GLMNET \citep{glmnetpack}, two of the most highly used algorithms for high-dimensional variable selection for survival data. We used the published R packages of those two methods to run the simulations. We also performed a comparison with a case when pMOM priors are used as the prior for nonzero coefficients instead of piMOM priors.

The \emph{``Aim''} of the simulation study was to compare the performance of our method with the other two methods with respect to the correlation structure between covariates in the design matrix. More specifically, we reported three different simulation settings that consider different combinations of correlation structure, true model size and the magnitude of true coefficients. This was the basis of our \emph{``Data generating mechanism''}. The correlation structure used in those settings are similar to the simulations reported in \citet{fansurvisis}.

%We first examined the six simulation settings described in \citet{fansurvisis}. These settings consider different aspects of variable selection with respect to the correlation between true covariates and the magnitude of true coefficients. {\color{red} Here, we use the correlation structure in the two of the hardest settings, which \citet{fansurvisis} called \emph{equi-correlated covariates with correlation = 0.5} and \emph{two very hard variables to be selected.} In total, we report 3 simulation cases with the following details.} Survival times a We refer to these settings as Case 1 and 2, respectively. We also added a third case in which the survival times are sampled from a Weibull distribution with uniform censoring.

For Case 1, $X_1,\ldots,X_p$ were multivariate Gaussian random variables with mean $0$ and marginal variance of $1$. The correlation structure was $\text{corr}(X_i,X_5)=0$ for all $i\neq{4,5}$, $\text{corr}(X_5,X_4)=1/\sqrt{2}$ and $\text{corr}(X_i,X_j)=0.5$ for $i,j \in \{1,\ldots,p\}\setminus\{4,5\}$. The size of the true model was 5 with nonzero regression coefficients $\beta_1= -1.5389, \beta_2=0.6839, \beta_3=-0.8498, \beta_4=-1.2716, \beta_5=-1.1045$ and $\beta_i=0$ for $i>5$. %The nonzero coefficients were chosen randomly, and were generated as $Z\cdot U$ with $Z\sim \text{Unif}(0,2)$ and $U=1$ or $U=-1$ with probability 0.5.%
The number of observations and covariates were $n=400$ and $p=1000$. The censoring rate for this case was approximately $27.6\%$. The survival and censoring times are both sampled from an exponential distribution. The rate parameter for the distribution of censoring times was set to $0.1$.

%For Case 1, $X_1,\ldots,X_p$ are multivariate Gaussian random variables with mean $0$ and marginal variance of $1$. The correlation structure between variables is $\text{corr}(X_i,X_j)=0.5$. The size of the true model is six with non-zero regression coefficients $\beta_1= -1.5140, \beta_2=1.2799, \beta_3=-1.5307, \beta_4=1.5164, \beta_5=-1.3020, \beta_6=1.5833$, and $\beta_i=0$ for $i>6$. The number of observations and covariates are $n=400$ and $p=1000$. The censoring rate for this case is $23\%$.

%For Case 2, $X_1,\ldots,X_p$ are multivariate Gaussian random variables with mean $0$ and marginal variance of $1$. {\color{red} The correlation structure is $\text{corr}(X_i,X_5)=0$ for all $i\neq{4,5}$, $\text{corr}(X_5,X_4)=1/\sqrt{2}$, and $\text{corr}(X_i,X_j)=0.5$ for $i,j \in \{1,\ldots,p\}\setminus\{4,5\}$.} The size of the true model is five with non-zero regression coefficients $\beta_1= 4, \beta_2=4, \beta_3=4, \beta_4=-6\sqrt{2}, \beta_5=4/3$ and $\beta_i=0$ for $i>5$. The censoring rate for this case is $38\%$. Similar to the previous case, the number of observations and covariates are $n=400$ and $p=1000$. In both Case 1 and Case 2 the survival times and censoring times are sampled from an exponential distribution. The rate parameter for the distribution of censoring times is set to $0.1$.

For Case 2, $X_1,\ldots,X_p$ were multivariate Gaussian random variables with mean $0$ and marginal variance of $1$. The correlation structure between variables was $\text{corr}(X_i,X_j)=0.5; i\neq j$. The size of the true model was 6 with nonzero regression coefficients $\beta_1= 1.1201, \beta_2=0.8322, \beta_3=-1.9620, \beta_4=-1.7639, \beta_5=1.6782, \beta_6=1.8995$, and $\beta_i=0$ for $i>6$. %As in Case 1, the nonzero coefficients were chosen randomly, and were generated as $Z\cdot U$ with $Z\sim\text{Unif}(0,2)$ and $U=1$ or $U=-1$ with probability 0.5.%
The number of observations and covariates were $n=400$ and $p=1000$.  In this case the survival times were sampled from a Weibull distribution with rate parameter $\lambda=0.1$ and shape parameter $k=15$. The censoring times were sampled uniformly from $[0,8]$, and the resulting censoring rate for this case was approximately $14.8\%$.

\sloppy For Case 3 the design matrix and correlation structure between variables was the same as Case 2, where $\text{corr}(X_i,X_j)=0.5,\; i\neq j$. The size of the true model was 20 with nonzero regression coefficients $(\beta_1,\ldots,\beta_{20})$ equal to (-1.6802, -1.2483,  2.9430, -2.6458, -2.5173, -2.8493, -2.0070, -1.5931, 0.8800, -0.9387,  1.6599, -2.9288, -1.2495, -2.6298, -2.3434, 1.9075, -1.1044, -0.7873, 2.6722, -0.6340), %
%$\beta_1= -1.6702, \beta_2=-1.7122, \beta_3=0.2313, \beta_4=-1.3798, \beta_5=0.9305, \beta_6=-1.3835, \beta_7= -1.8575, \beta_8=1.0676, \beta_9=1.2013, \beta_{10}=0.5116, \beta_{11}=0.9871, \beta_{12}=-0.7680, \beta_{13}= 0.5159, \beta_{14}=-0.2649, \beta_{15}=-1.7482, \beta_{16}=-1.6041, \beta_{17}=1.4234, \beta_{18}=-1.2884, \beta_{19}=-1.1299, \beta_{20}=1.2569$,%
and $\beta_i=0$ for $i>20$. %Like the other two cases, the nonzero coefficients were chosen randomly but with larger upper limit on the uniform distribution. They were generated as $Z\cdot U$ with $Z\sim\text{Unif}(0,3)$ and $U=1$ or $U=-1$ with probability 0.5.%
The number of observations and covariates were $n=400$ and $p=1000$.  The censoring rate for this case was approximately $34.1\%$. The survival and censoring times were both sampled from an exponential distribution. The rate parameter for the distribution of censoring times was set to $0.1$.

Each simulation case was then repeated 50 times, $n_{iter}=50$, and each time with different random seed numbers in order to generate different datasets.

The primary targets of our simulation study, or the \emph{``Estimands,''} according to \citet{morris}, were identifying the true model as well as estimating the vector of coefficients of the true model. Accordingly, we reported four different quantities as \emph{``Performance measures''} for those estimands. The first two quantities are the mean $l_1$ norm of the error in estimating the vector of coefficients and the mean squared error (MSE). The mean $l_1$ norm was computed as $\frac{1}{n_{iter}}\sum_{i=1}^p |\hat{\beta}_i-\beta_i|$, and the MSE was computed as $\frac{1}{n_{iter}}\sum_{i=1}^p (\hat{\beta}_i-\beta_i)^2$.  The third quantity is the mean model size of the selected models and was denoted by MMS. MTP and MFP denote mean false positive and mean true positive values for each algorithm. %The last outcome is the proportion of times that the algorithm selected the true model (without any false positives {\color{red} or false ngatives}). This proportion is denoted by $P$.
Formal definitions of MFP, MTP are provided in Section 3 of the \ref{suppA} \citep{supplement}. 

Table \ref{c4:simres} compares the performance of our method, BVSNLP, the default settings of ISIS-SCAD and GLMNET algorithms. The $\lambda$ parameter in GLMNET was picked by cross validation. Table \ref{mser} compares the Monte Carlo standard errors \citep{morris} of the MSEs for all three methods. %The LASSO method is not in listed in that table because it takes several days to complete a single repetition of any of these simulation cases \citep{fansurvisis}.

\begin{table}[!t]
	\centering
	\caption{Comparison between BVSNLP, ISIS-SCAD and GLMNET for simulation Cases 1, 2, and 3 	with $n=400$ and $p=1000$.}\label{c4:simres}
	\begin{tabular}{l|ccc}
		\hline
		\rule{0pt}{3ex}& \textbf{BVSNLP} & \textbf{ISIS-SCAD} & \textbf{GLMNET} \\\hline
		
		%		\rule{0pt}{3ex}\textbf{Case 1:} &&& \\ % The usual case in Fan paper.
		%		Mean $l_2$ norm  & 0.053 & 0.225 & 1.072  \\
		%		Mean $l_1$ norm  & 0.449 & 0.465 &  3.924 \\
		%		MMS & 6 & 6 & 55.36  \\
		%		MTP & 6 & 6 & 6 \\		
		%		MFP & 0 & 0 & 49.36 \\
		%		P & 1 & 1 & 0\\ \hhline{====}
		
		\rule{0pt}{3ex}\textbf{Case 1:} &&& \\ % The hardest case in Fan paper.
		MSE  & 0.141 & 0.792 & 1.441  \\
		Mean $l_1$ norm  & 0.488 & 2.200 &  4.072 \\
		MMS & 4.96 & 8.84 & 51.46  \\
		MTP & 4.92 & 4.62 & 4.00 \\
		MFP & 0.04 & 4.22 & 47.46 \\ \hhline{====}
		%P & 1 & 0.98 & 0\\ \hhline{====}
		
		\rule{0pt}{3ex}\textbf{Case 2:} &&& \\ % The weibull as recommended by Leo Held.
		MSE  & 0.141 & 0.792 & 1.441 \\
		Mean $l_1$ norm  & 0.505 & 0.552 &  3.891 \\
		MMS & 6 & 5.94 & 50.94  \\
		MTP & 6 & 5.88 & 5.92 \\
		MFP & 0 & 0.06 & 45.02 \\ \hhline{====}
		%P & 1 & 1 & 0\\ \hhline{====}
		
		\rule{0pt}{3ex}\textbf{Case 3:} &&& \\ % The 20 variable case that has SISGLM fail, with no correlation between columns.
		MSE  & 0.602 & 5.287 & 4.701  \\
		Mean $l_1$ norm  & 2.680 & 22.962 &  22.824 \\
		MMS & 20.08 & 14.76 & 105.62 \\
		MTP & 19.94 & 12.80 & 19.96 \\
		MFP & 0.14 & 1.96 & 85.66 \\\hline
		%P & 0 & 0 & 0\\\hline		
	\end{tabular}
	
\end{table}

\begin{table}[!hb]
	\centering
	\caption{Monte Carlo Standard Errors for the MSE of the coefficient vector for all three methods.}\label{mser}
	\begin{tabular}{l|ccc}
		\hline
		\rule{0pt}{3ex}& \textbf{Case 1} & \textbf{Case 2} & \textbf{Case 3} \\\hline
		\rule{0pt}{3ex}\textbf{BVSNLP}& 0.064 & 0.008 & 0.065\\\hline		
		\rule{0pt}{3ex}\textbf{ISIS-SCAD}& 0.098 & 0.015 & 0.142\\\hline		
		\rule{0pt}{3ex}\textbf{GLMNET}& 0.007 & 0.024 & 0.055\\\hline										
	\end{tabular}
\end{table}

In the S5 algorithm 30 iterations were used within each temperature. The parameter $d$ was chosen as $2\ceil[\big]{\log(p)}$. As described in Section \ref{c4:as5}, $d$ represents the number of candidate covariates that were added to the current model to make the addition set, $\Gamma^+$. Each S5 algorithm was run in parallel on 120 CPUs for all three simulation cases. The beta-binomial prior was imposed on the model space with $a=1, b=p-a$. The hyperparameters of the piMOM prior were selected using the algorithm discussed in Section \ref{hypersel} with $\alpha=0.8$ for all three simulation cases, imposed as the prior mode.

Finally, the average runtimes of the BVSNLP algorithm for the three simulation cases were 29, 20.23 and 27.99 seconds, respectively.  %entire simulation is summarized in Table \ref{c4:simtime}. 

%\begin{table}[!b]
%	\centering
%	\caption{Average BVSNLP run time over 50 iterations in each simulation case using 120 CPUs.}\label{c4:simtime}
%	\begin{tabular}{l|ccc}
%		\hline
%		\rule{0pt}{3ex}& \textbf{Case 1} & \textbf{Case 2} & \textbf{Case 3} \\\hline
%		\rule{0pt}{3ex}Run time (seconds)& 29.00 & 20.23 & 27.99\\\hline		
%	\end{tabular}
%\end{table}

As demonstrated in Table \ref{c4:simres}, our method performs better than the other two methods according to all selected metrics, regardless of the size of the true model. The difference between BVSNLP and ISIS-SCAD is best illustrated as the size of the true model increases. GLMNET has significantly higher mean false positive rates than the other two methods.

\begin{figure}[!t]
	\centering
	\includegraphics[width=80mm]{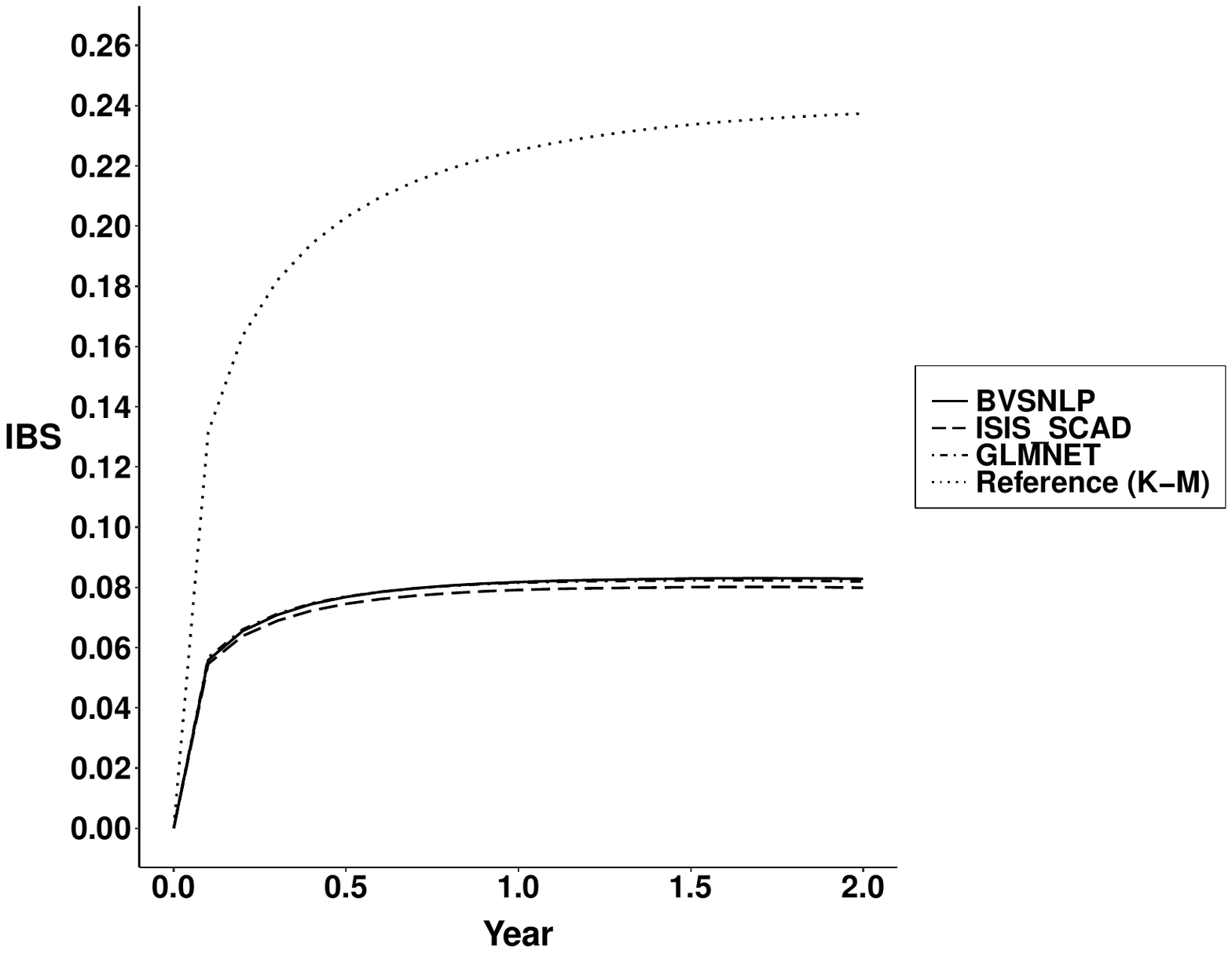}
	\caption{Average IBS for all methods in simulation case 1.}\label{ibssim1}
\end{figure}

\begin{figure}[!t]
	\centering
	\begin{subfigure}{.5\textwidth}
		\centering
		\includegraphics[width=1\linewidth]{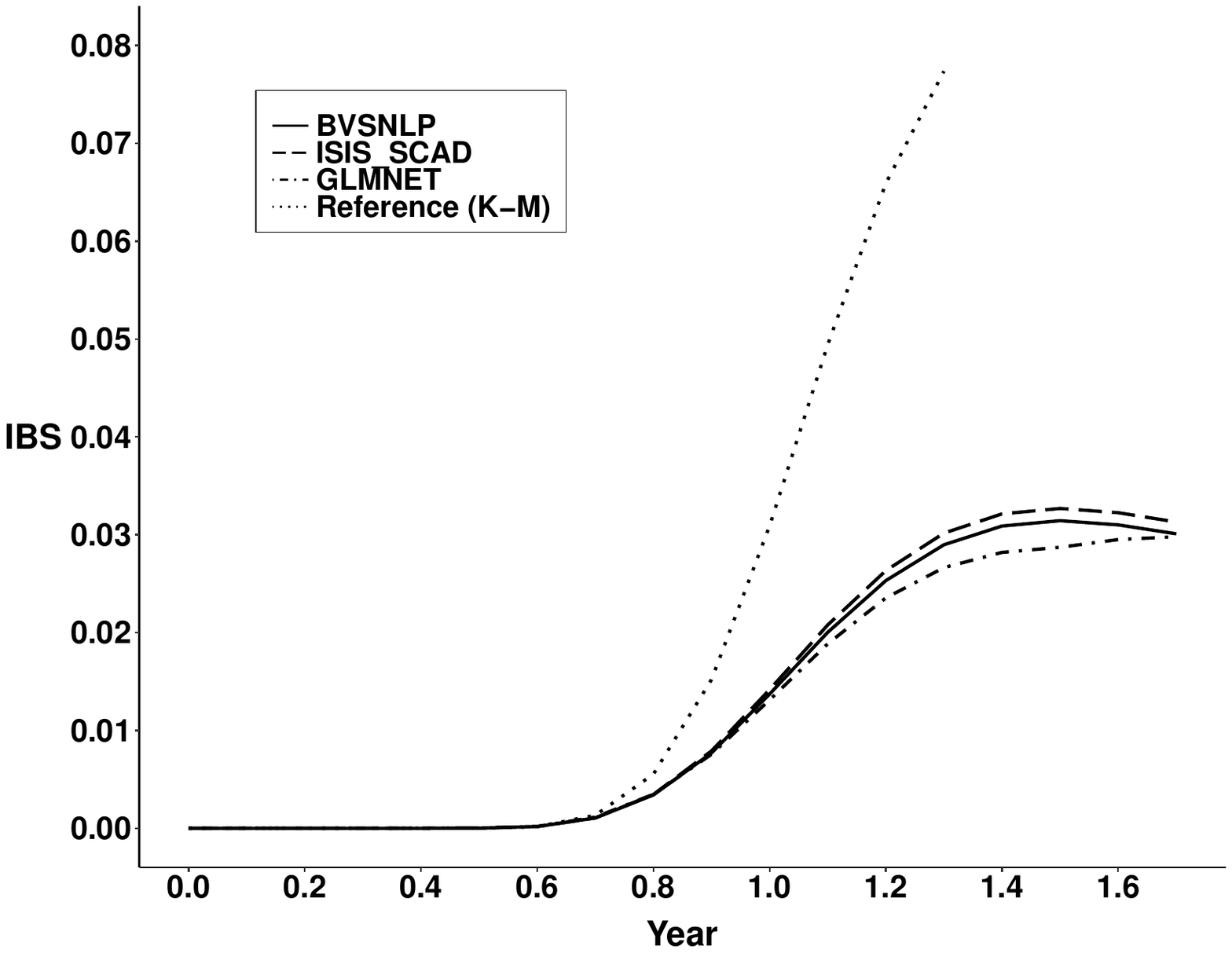}
		\caption{Simulation case 2.}
		\label{ibssim2}
	\end{subfigure}%
	\begin{subfigure}{.5\textwidth}
		\centering
		\includegraphics[width=1\linewidth]{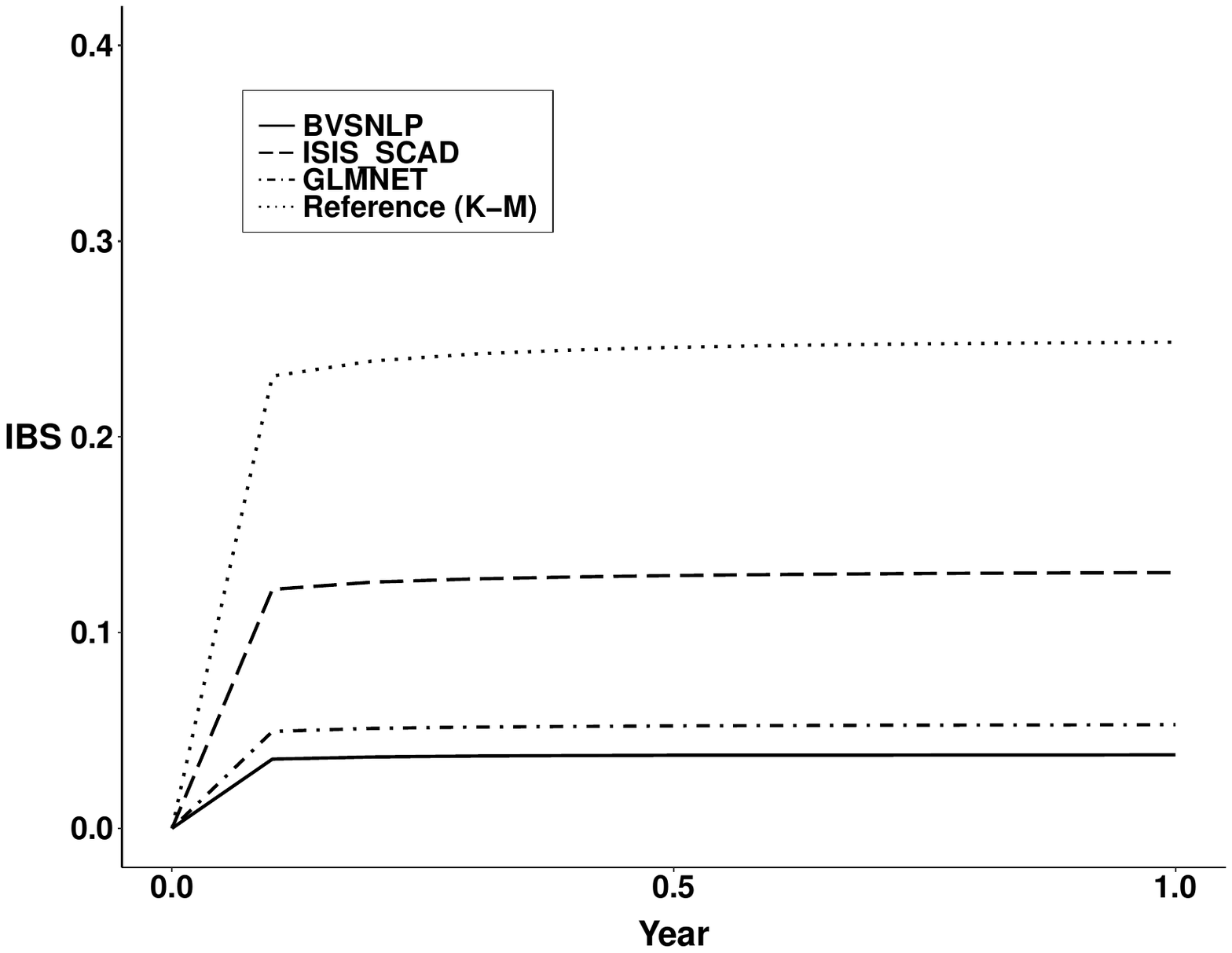}
		\caption{Simulation case 3.}
		\label{ibssim3}
	\end{subfigure}
	\caption{IBS plots for simulation cases 2 and 3.}
	\label{ibsc2c3}
\end{figure}

Figures \ref{ibssim1},  \ref{ibssim2} and \ref{ibssim3} compare the average IBS over 50 iterations between the methods discussed above.  IBS is computed using the R package \verb|pec| \citep{pecpack} based on a five-fold cross-validation. A benchmark model based on Kaplan--Meier estimate, which includes no covariates, is also added to the figures as a reference for the comparison. The average c-index measures for all the methods are also reported in Table \ref{cindsim}. The c-index measures are computed based on the method discussed in \citet{cindexbook}, using the \verb|dynpred| package in R. Because a new dataset was created at each iteration, it was not possible to get the average prediction errors, due to the fact that the times points where prediction errors change were different for different data sets. 

As shown in the IBS plots, all three methods perform better than the reference. BVSNLP and ISIS-SCAD have a very similar performance. For Case 3, where the true model has 20 covariates, BVSNLP outperforms the other two methods, whereas in Case 2, GLMNET has the best performance. The c-index is similar for all methods and seems to provide a smaller penalty for model size.  This feature of the c-index is discussed further in Section \ref{c4:discuss}.

\begin{table}[!hb]
	\centering
	\caption{Average c-index measures over 50 iterations in each simulation case.}\label{cindsim}
	\begin{tabular}{l|ccc}
		\hline
		\rule{0pt}{3ex}& \textbf{Case 1} & \textbf{Case 2} & \textbf{Case 3} \\\hline
		\rule{0pt}{3ex}\textbf{BVSNLP}& 0.890 & 0.881 & 0.960\\\hline
		\rule{0pt}{3ex} \textbf{ISIS-SCAD} & 0.895 & 0.876 & 0.841\\\hline		
		\rule{0pt}{3ex}\textbf{GLMNET}& 0.911 & 0.908 & 0.970\\\hline				
	\end{tabular}
\end{table}

\section{Application to Real Data}\label{c4:resultreal}
We applied our method to selected genes associated with patient survival times for two common cancer types using datasets from The Cancer Genome Atlas (TCGA) projects: kidney renal clear cell carcinoma (KIRC) \citep{naturerenal} and kidney renal papillary cell carcinoma (KIRP) \citep{kirp}. We compared the performance of our algorithm to ISIS-SCAD \citep{fansurvisis}, GLMNET \citep{glmnetpack} and Stability Selection \citep{stability}.  Stability Selection was combined with a high-dimensional selection algorithm, such as GLMNET, and selects the most stable features for a given level of Type I error. To run the Stability Selection method, we used the \verb|c060| R package \citep{c060} and the recommended values for function arguments.

We included patients' ``Age, Gender'' and a clinical stage variable, ``Stage'', in the design matrix. On the advice of a clinician, the ``Stage'' variable  was developed by combining the histological stage, pathological stage and clinical stage into one variable that is a summary of how advanced each subject's cancer was when the tissue sample was taken.``Stage'', like ``Gender'', is a categorical variable but with three levels, where ``Stage $i$'' represents the $i^{th}$ class of that variable; ``Stage 3'' represents the most advanced stage. To remove stromal contaminations from the gene expression data, the DeMixT algorithm \citep{demix2} was performed on the design matrix and the tumor-specific expression data were used in the analyses for all algorithms. %We compared our method to ISIS-SCAD, GLMNET and Stability Selection \citep{stability}. Stability Selection is combined with a high dimensional selection algorithm such as glmnet and selects the most stable features for a given level of Type I error.

The predictive performance was measured by a time-dependent AUC, as discussed in Section \ref{predictive}, based on a five-fold cross-validation. The observations in each fold were randomly chosen under a constraint which balanced censoring rate between folds. The AUC values were computed for the test set using the model that was obtained by performing variable selection on the training set. The selected covariates for each cancer type were also compared. For our method, we report the covariates associated with the highest posterior probability model. The hyperparameter $\tau$ of the piMOM prior was selected using the algorithm in Section \ref{hypersel} with $\alpha=0.1$ as the mode of the piMOM prior. This is our choice of $\alpha$ for real datasets. The results for each cancer type are discussed in separate sections below. Note that GLMNET has a random output when the hyperparameter is selected by cross-validation. As a result, based on the recommendation of the inventors of that algorithm, we ran it 100 times for each fold and took the average of results as the outcome for that fold. 

We treated categorical variables ``Stage'' and ``Gender'', as well as the continuous variable ``Age'', as fixed covariates in our model. However, available ISIS-SCAD and Stability Selection software packages are not able to fix preselected covariates to be included in all models. For this reason, dummy variables associated to ``Stage'' and ``Gender''  were manually added to the design matrix and were subject to the selection procedure for those methods. 

\subsection{Kidney Renal Clear Cell Carcinoma (KIRC)}
After removing covariates with missing expressions and observations with missing survival times, the KIRC dataset \citep{naturerenal} contains 490 observations with 13,267 covariates, . The censoring rate for this dataset is 66.94\%. Table \ref{kirc_genes} shows the covariates selected by each method. As mentioned previously, GLMNET produces random outputs at each run, and therefore, for this table, only the output for one of the runs are indicated; other runs produced a similar number of selected covariates.

\begin{table}[!t]
	\centering
	\caption{Selected genes and covariates for KIRC across different variable selection algorithms}\label{kirc_genes}		
	\begin{tabular}{l|lll}
		\hline
		\rule{0pt}{3ex}\multirow{2}{*}{\textbf{BVSNLP}} & Age & Gender  & Stage \\ 
		&SUDS3 & AR & \\\hline
		\rule{0pt}{3ex}\multirow{5}{*}{\textbf{ISIS-SCAD}} & Stage 3 & AR & Age  \\
		&HEBP1&ATP2C1&GADD45A\\
		&MTERF2&ADGRL3&GPSM1\\
		&SERPINI1&SP6&ZNF815P\\
		&INAFM2 && \\\hline
		\rule{0pt}{3ex}\multirow{8}{*}{\textbf{GLMNET}} & Stage & AR & Age\\
		& HEBP1 & SEC61A2 & TRMT6 \\
		& PCBP4 & FAHD2A & MCM8\\
		& E2F5 & SLC5A6 & NARF\\
		& RAB28 & DONSON & GPSM1\\
		& HACD1 & MARS & FASN\\
		& TRAIP & RPL17P50 & SLC26A6\\
		& GPR162 & INAFM2 & ACACA\\\hline
		\rule{0pt}{3ex}\multirow{2}{*}{\textbf{Stability Selection}} & Stage 3 & AR & Age \\
		&INAFM2 && 
	\end{tabular}
\end{table}

In addition to the categorical covariate ``Stage'', BVSNLP selects ``AR'' and ``SUDS3'' in the HPPM as the most significant covariates in the design matrix. The posterior inclusion probabilities for ``AR'' and ``SUDS3'' are $0.80$ and $0.08$, respectively. The ``Age'', ``Gender'',  and ``Stage'' were fixed in all models and thus were selected with probability $1$. The MAP estimates for the coefficients of ``Age, Gender Male, Stage 2, Stage 3, AR'' and ``SUDS3'' were $0.33, -0.11, 0.45, 1.61, -0.60 $ and $0.36$, respectively. These coefficients indicate that patients with the most advanced stages of cancer had the poorest survival rates, and that a patient with a tumor sample characterized as advanced has a hazard rate that was $\exp(1.61) \approx 5$ times higher than a patient with tumor sample characterized as localized, when all other covariates were the same. These coefficients also show that the hazard rate in females is 1.12 times that in males, and age has an unfavorable impact on the hazard rate, as expected. Moreover, the negative sign for the ``AR'' gene indicates it has a favorable impact on survival for KIRC. ``AR'', the Androgen Receptor gene, functions as a steroid-hormone activated transcription factor. It has been well documented that ``AR'' promotes the progression of renal cell carcinoma (RCC) through hypoxia-inducible factors HIF-2$\alpha$ and vascular endothelial growth factor regulation \citep{ar1}. The favorable impact of the ``AR'' gene was also studied by \citet{arfav} in bladder cancer. ``SUDS3'' is a regulatory protein that is part of the SIN3A corepressor complex component that potentially has a role in tumor suppressor pathways through regulation of apoptosis. There was previous evidence of the down-regulation of the SIN3A gene in tumorigenesis of lung cancer \citep{suds3}. 

%We also performed variable selection by searching for all first order interaction terms with AR. The interaction of three genes, ARHGEF10, TNIP1 and LINC00894 with AR were selected by our algorithm. ARHGEF10 was reported to locate in a candidate tumor suppressor gene region \citep{arh}. TNIP1 plays a role in autoimmunity and tissue homeostasis through the regulation of nuclear factor kappa-B activation and has been reported before to associate with various cancers. 

%We did further analysis and included the interaction terms between selected genes. However, none of the coefficients for interaction terms were significantly different from zero based on their Wald statistic value and therefore, the interaction terms are not discussed in our analyses.

%The MLE of coefficients are summarized in Table \ref{kirc_coef}. SUDS3 and AR genes are the most significant of all whereas none of the interaction terms were significant in predicting the survival. AR has a favorable impact on survival where a unit increase in its expression doubles the survival rate. In contrast, SUDS3 has an unfavorable impact on survival where an increase in that gene causes almost $50\%$ increase in hazard rate. {\color{red}some literature to back this finding.} As expected, the stage 4 of the cancer also has an unfavorable effect on the survival.

It is noteworthy that the algorithm selected the same highest posterior probability model for different values of the hyperparameter $\tau$ in the range $[0.01,0.9]$. This shows the robustness of the proposed variable selection algorithm to the choice of hyperparameter $\tau$ for a range of plausible values.

%\begin{table}[h]
%	\centering
%	\caption{Coefficients of selected interaction terms and the main effects}\label{kirc_coef}		
%	\begin{tabular}{l|c|c}
%		\hline
%		Covariates & Coefficient & $\exp$(Coefficient)\\ \hline
%		SUDS3 & 0.323 & 1.381 \\
%		AR & -0.699 &  0.497  \\		
%		Stage 4 & 0.346 & 1.413 \\
%		Stage4:SUDS3 & 0.188 & 1.206 \\
%		Stage4:AR & 0.015 & 1.016 \\
%		SUDS3:AR & 0.020 & 1.020 \\\hline	
%	\end{tabular}
%\end{table}

For this particular run of GLMNET, a much larger model was selected with 24 variables including two of the variables reported by BVSNLP. ISIS-SCAD selected 13 covariates, which included the four covariates that were selected by the Stability Selection method. ``AR'' and the last level of ``Stage'' are the common covariates among all methods.

\begin{figure}[!t]
	\centering
	\includegraphics[width=80mm]{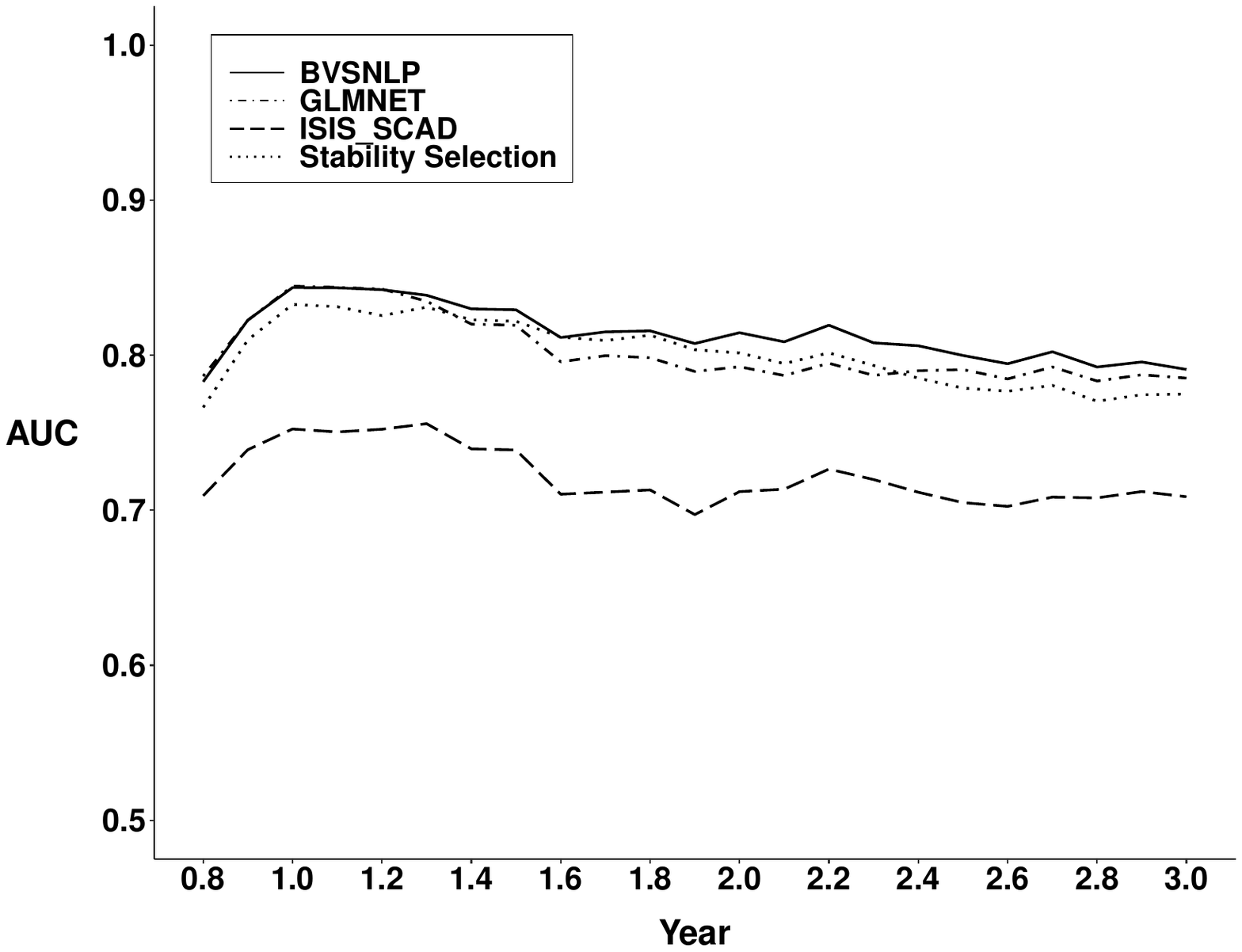}
	\caption{Average AUC of different variable selection methods based on a five fold cross-validation for KIRC dataset.}\label{kircauc}
\end{figure}

The time dependent AUC plot for all four methods, obtained by performing a five-fold cross validation, is depicted in Figure \ref{kircauc}. BVSNLP has slightly better predictive accuracy than GLMNET and Stability Selection.  However, it achieves this accuracy with a much sparser model. We investigated the covariates that were selected by each of those algorithms in all five folds and found that BVSNLP, in addition to those fixed covariates, selects only 10 unique genes in total, where `AR' is selected in three of the five folds. %In other words, two out of three covariates that are selected using the full data were selected in all five training datasets in cross validation.

GLMNET selected 160 different covariates across all five folds. Only five out of 24 selected covariates in Table \ref{kirc_genes} were selected in all five training datasets in cross-validation. Those include ``Age, Stage'' and ``AR''. GLMNET was run 100 times for each fold. ISIS-SCAD selected 45 different covariates, and only ``Stage 3'' was selected in all training datasets in cross-validation. The Stability Selection method selected sparser models compared to ISIS-SCAD and GLMNET by selecting 13 different covariates. It picked only ``Age'' and ``Stage 3'' in all five folds.

\begin{figure}[!t]
	\centering
	\begin{subfigure}{.5\textwidth}
		\centering
		\includegraphics[width=1\linewidth]{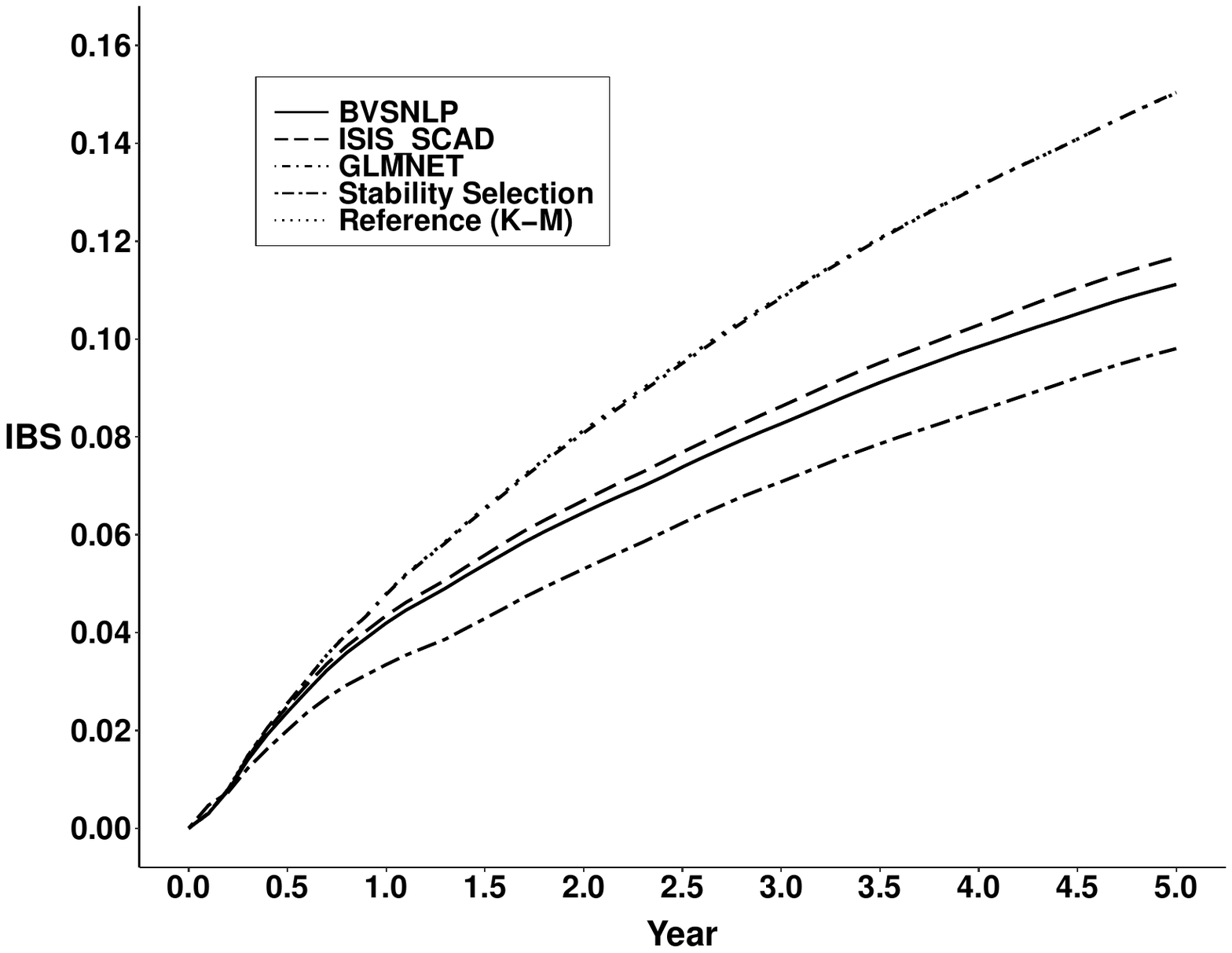}
		\caption{IBS}
		\label{ibskirc}
	\end{subfigure}%
	\begin{subfigure}{.5\textwidth}
		\centering
		\includegraphics[width=1\linewidth]{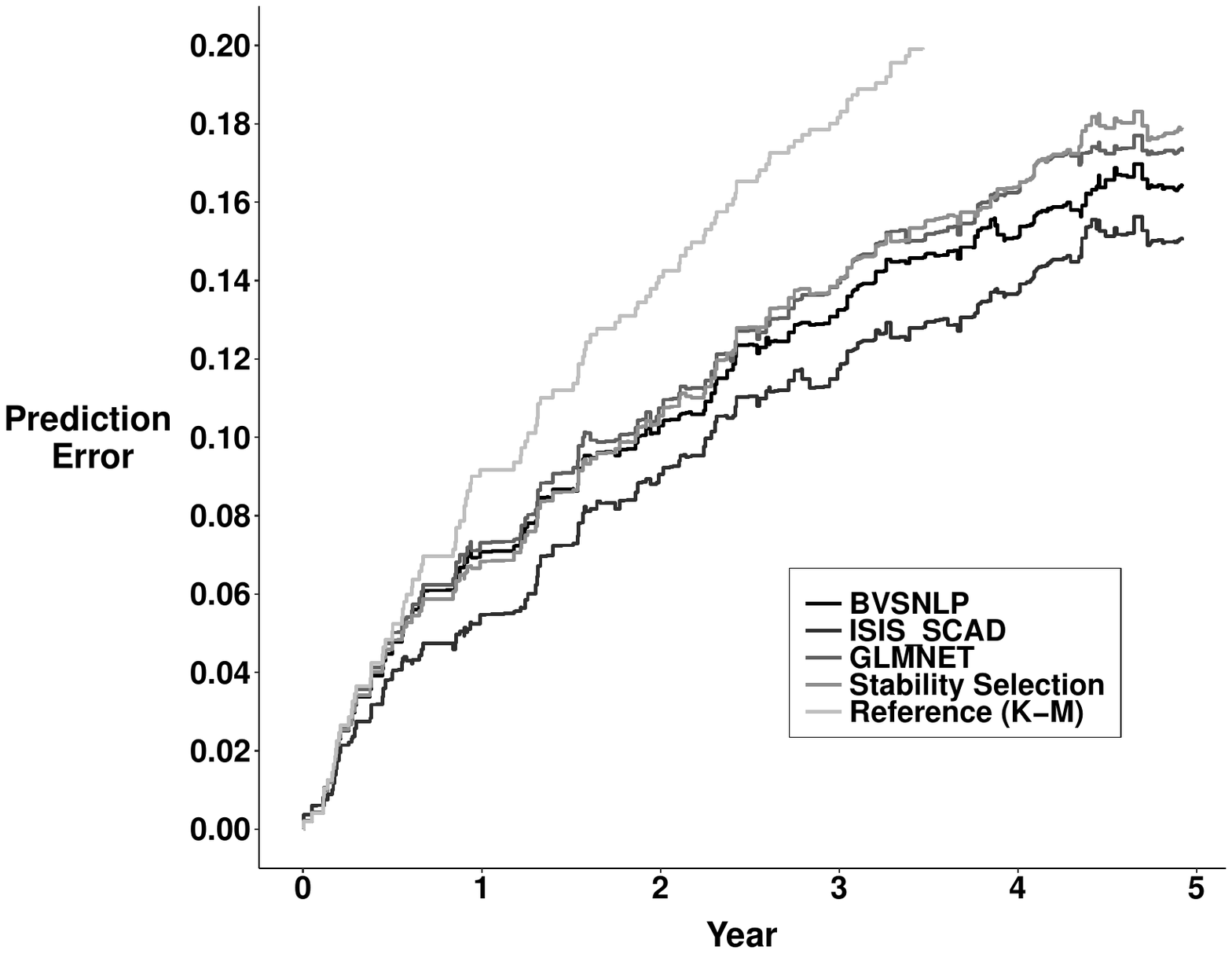}
		\caption{Prediction Error}
		\label{perrkirc}
	\end{subfigure}
	\caption{IBS and prediction error of BVSNLP for the KIRC dataset.}
	\label{perribskirc}
\end{figure}

Figures \ref{ibskirc} and \ref{perrkirc} compare IBS and prediction error curves, respectively, between different methods for the KIRC dataset. These two measures were computed based on a five-fold cross-validation. Computation of IBS and prediction error were done using the R package \verb|pec| \citep{pecpack}. A benchmark model based on the Kaplan--Meier estimate, which includes no covariates, was also added to the figures as a reference for the comparison. The c-index measures are also reported in Table \ref{cindkirc}. The c-index was computed as it was in Section \ref{c4:resultsimul} using the \verb|dynpred| package in R.

%\begin{figure}[!t]
%		\centering
%		\includegraphics[width=80mm]{ibs_kirc}
%		\caption{IBS comparison between all methods for the KIRC dataset.}\label{ibskirc}
%\end{figure}

\begin{table}[!hb]
	\centering
	\caption{Average c-index measure of different methods for the KIRC dataset.}\label{cindkirc}
	\begin{tabular}{l|cccc}
		\hline
		\rule{0pt}{3ex}& \textbf{BVSNLP} & \textbf{GLMNET} & \textbf{ISIS-SCAD} &\textbf{Stability Selection} \\\hline
		\rule{0pt}{3ex}c-index measure& 0.804 & 0.816 & 0.846 & 0.797\\\hline		
	\end{tabular}
\end{table}

%\begin{figure}[!hb]
%		\centering
%		\includegraphics[width=80mm]{perr_kirc}
%		\caption{Comparison of prediction errors between all methods for the KIRC dataset.}\label{perrkirc}
%\end{figure}

GLMNET has almost the same IBS curve as the reference Kaplan--Meier curve. BVSNLP outperforms ISIS-SCAD, and Stability selection has the best IBS performance among all. For prediction error curves, BVSNLP is second to ISIS-SCAD, and GLMNET and Stability Selection have almost the same performance. A different behavior can be seen for c-index measures where GLMNET and ISIS-SCAD have higher c-indices than BVSNLP. %It shows that these measures are not consistent in showing predictive performance of the methods. This is discussed further in Section \ref{c4:discuss}.

BVSNLP was run on 120 CPUs, Stability Selection was run on four CPUs, while GLMNET and ISIS-SCAD were run on a single CPU. The average runtime for different methods in each fold of the cross validation was 6.4, 180, 5, and 1.3 minutes for BVSNLP, GLMNET, ISIS-SCAD and Stability Selection.

%\begin{table}[!hb]
%		\centering
%		\caption{Average run time for different methods in each fold of the cross validation for KIRC data set.}\label{c4:avrt}
%		\begin{tabular}{l|cccc}
%			\hline
%			\rule{0pt}{3ex}& \textbf{BVSNLP} & \textbf{GLMNET} & \textbf{ISIS-SCAD} &\textbf{Stability Selection} \\\hline
%			\rule{0pt}{3ex}Run time (minutes)& 6.4 & 180 & 5.0 & 1.3\\\hline		
%		\end{tabular}
%\end{table}

In our previous study of binary outcomes using the same dataset \citep{imomlogit}, we performed hierarchical clustering on the deconvolved tumor-specific expression matrix and identified two clusters of patient	samples. We saw these two groups of patients present significantly different survival outcomes and, therefore, assigned good vs. bad survival to the groups. The dichotomization was based solely on the clustering results of deconvolved gene expression levels. Survival times and censoring did not play any role in that process. However, there was a loss of information in dichotomizing a survival dataset and analyzing it with logistic regression. Now, with BVSNLP, we are able to use the original survival time to event with censoring information. To compare the biological implications between the two analyses, we looked for known expression regulation networks between the gene sets found in the binary analysis, SAV1 and NUMBL, and the new genes found in this analysis, AR and SUDS3, using Pathway Studio\textsuperscript{\textregistered} \citep{nikitin, pstudio}. We found that the well-studied cancer genes TGFB1, BCL2, PPARG, NEDD4, and CTNNB1, and a regulatory microRNA, MIR21, constitute the shortest paths between SAV1 and AR. Similarly, we found cancer genes CDKN1A, WNT3A, two genes that determine cell fate (SOX17 (connected with CTNNB1) and NANOG) and PAX6 that regulates transcription, to constitute the shortest paths between NUMBL and SUDS3. These are depicted in Section 6 of the \ref{suppA} \citep{supplement} . These findings suggest a high biological consistency between our two analyses that used BVSNLP to select features for binary and survival outcomes.

\begin{figure}[!ht]
	\centering
	\includegraphics[width=80mm]{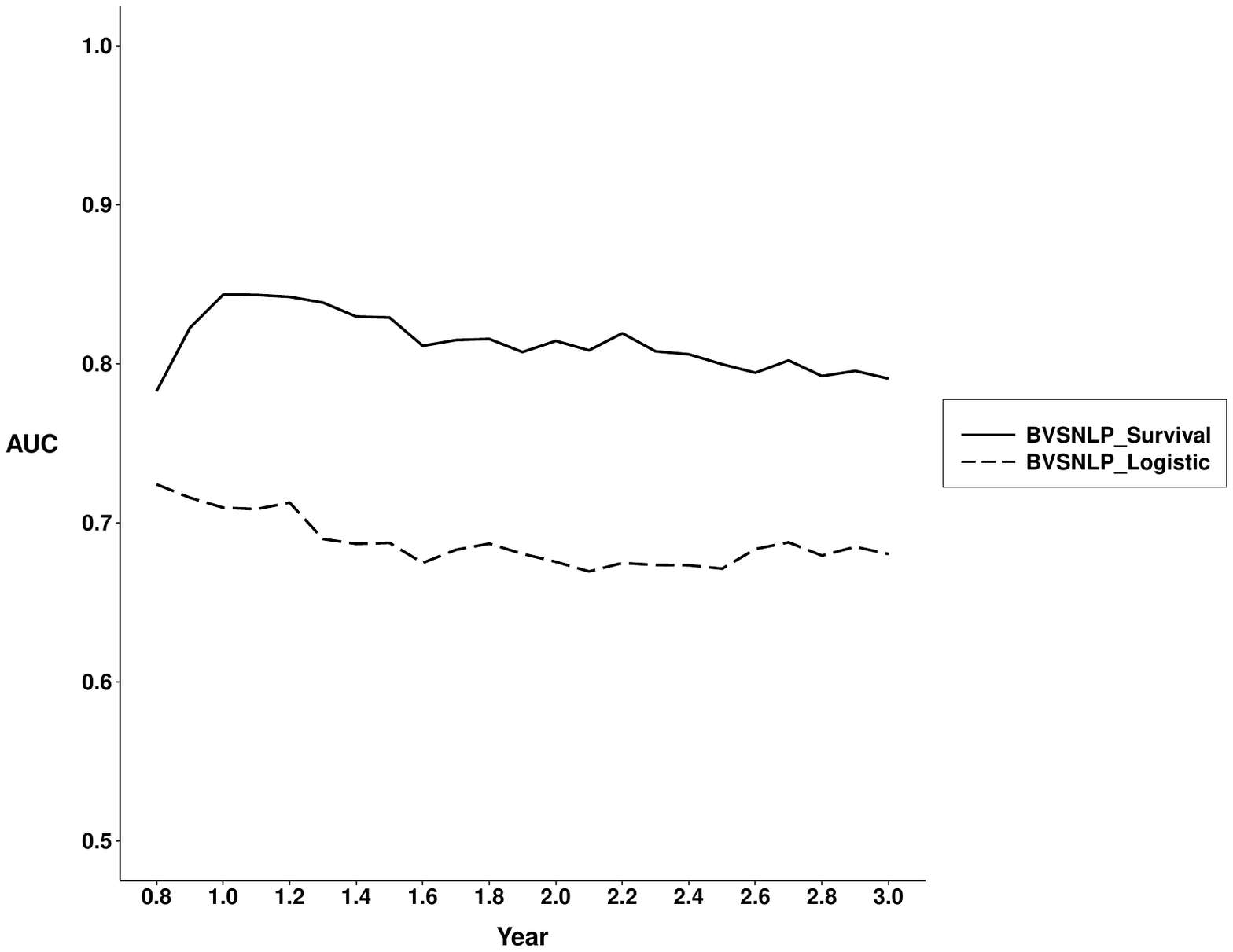}
	\caption{Comparison between BVSNLP model selection using survival  and dichotomized versions of the KIRC dataset.}\label{logitsurv}
\end{figure}

In summary, the binary model using SAV1 and NUMBL to predict overall survival of patients with kidney cancer is not as effective as the model using AR and SUDS3, as shown in Figure \ref{logitsurv}. Thus, although the findings of \citet{imomlogit} were biologically justified, some limitations were associated with those findings due to the information loss incurred by clustering and dichotomizing the data, and the BVSNLP survival model provides better insight on the genes associated with this cancer type.

\subsection{Kidney Renal Papillary Cell Carcinoma (KIRP)}
The KIRP dataset \citep{kirp} contains 244 samples with 13,335 covariates (after necessary data cleaning) and has a fairly high censoring rate of 85.7\%. The covariates selected by each method are summarized in Table \ref{kirp_genes}.

%ddBVSNLP selects the CDK1 gene as the most significant gene in this dataset. The posterior inclusion probability for `CDK1' is $0.314$, which is the highest posterior inclusion probability among all covariates. The total runtime of BVSNLP algorithm for this dataset was about 32 minutes. CDK1, the Cyclin Dependent Kinase 1, plays an important regulatory role in cell cycle control, hence in the progression of various cancer types. Multiple studies have suggested that CDK1 regulates the progression of renal cancer \citep{cdk11}, \citep{cdk12}, \citep{cdk13} but not specifically on KIRP.
%In our dataset, we obtained a coefficient of 1.0 for CDK1, suggesting an unfavorable prognosis for KIRP. Similar to the previous dataset, we tested BVSNLP for different choices of $\tau$ in the $(0,1)$ interval and the same model was selected for all of the cases. 

In addition to the fixed covariates ``Age, Gender'', and ``Stage'', BVSNLP selects the ``CDK1'' gene in the HPPM as the most significant covariate in the design matrix. The posterior inclusion probability for ``CDK1'' was $0.12$. The MAP estimates for the coefficients of ``Age, Gender Male, Stage 2, Stage 3'' and ``CDK1'' were $0.12, - 0.10, 0.11, 0.79$ and $1.13$, respectively. This shows that a unit increase in ``CDK1'' (Cyclin dependent kinase 1) gene expression increases the hazard rate by a factor of three for given values of the other covariates. CDK1 is a cell cycle regulator and has been reported previously as a prognostic marker gene for various cancer types. Many experimental studies have been performed to further understand the molecular mechanism behind the complex functions of CDK1 \citep{malumbres}. This is the first time, however, that CDK1 has been reported as a prognostic marker gene in human data for papillary renal cell carcinoma. As expected, patients at the most advanced stage cancer have a hazard rate that is 2.2 times higher than patients at a localized stage of cancer, given the values of all other covariates. As in the case of KIRC patients, age and male gender have unfavorable and favorable impacts on the hazard rate, respectively.

\begin{table}[!t]
	\centering	
	\caption{Selected covariates for KIRP across different variable selection algorithms}\label{kirp_genes}	
	\begin{tabular}{l|lll}
		\hline
		\rule{0pt}{3ex}\multirow{2}{*}{\textbf{BVSNLP}} & Age & Gender  & Stage \\ 
		& CDK1 && \\\hline
		\rule{0pt}{3ex}\textbf{ISIS-SCAD}&CDK1& COL6A1 &C19orf33 \\\hline
		\rule{0pt}{3ex}\textbf{GLMNET} & \multicolumn{3}{c}{\emph{No covariates were selected}}\\\hline
		\rule{0pt}{3ex}\textbf{Stability Selection} & Stage 3 & MTC02P12& RPL39P3
	\end{tabular}
\end{table}

%The maximum likelihood estimate of coefficient for the interaction between `Stage' and `TPX2' were not significantly different from zero based on its Wald statistic value.
\begin{figure}[!b]
	\centering
	\includegraphics[width=80mm]{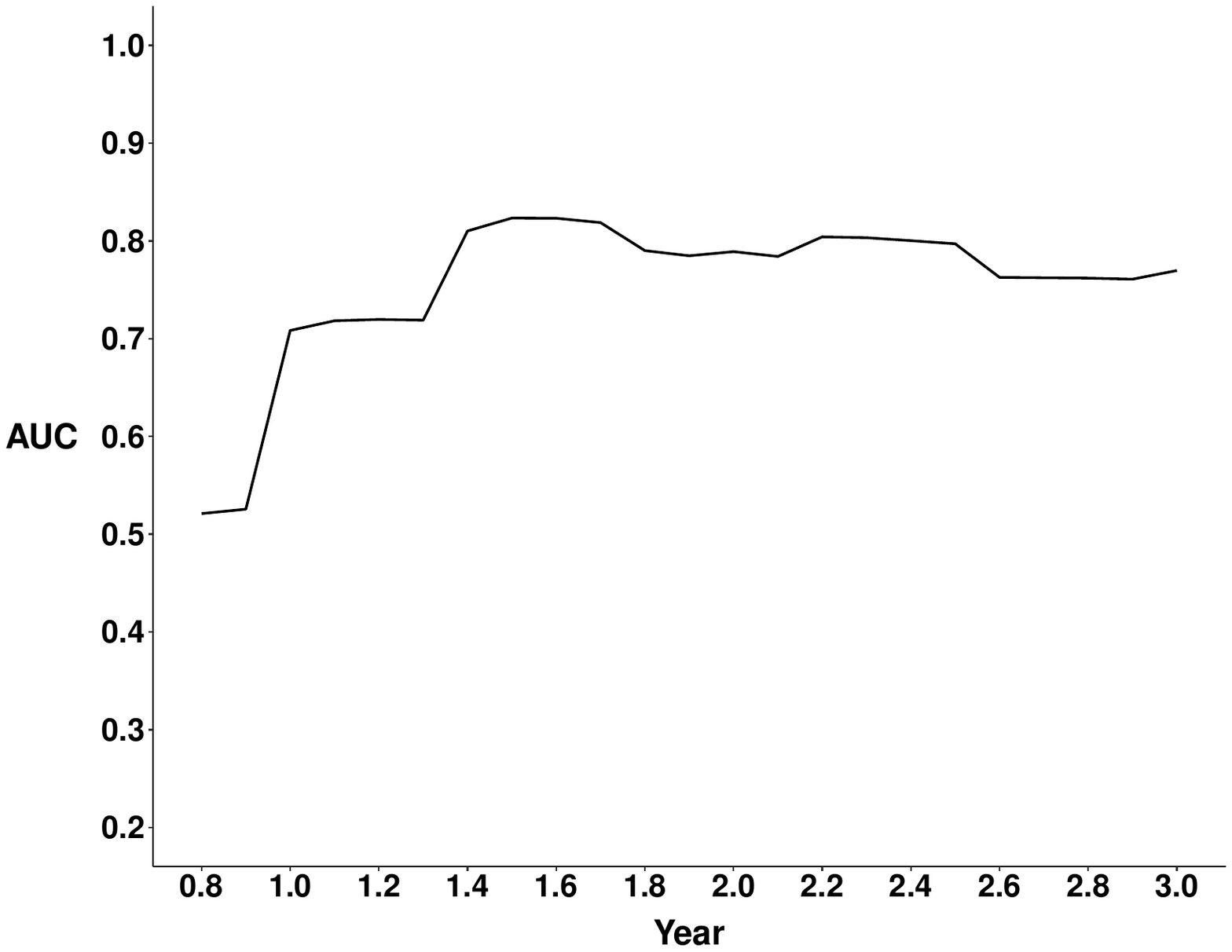}
	\caption{Average AUC of BVSNLP based on a five fold cross-validation for the KIRP dataset.}\label{kirpauc}
\end{figure}

Surprisingly, GLMNET does not select any covariates. Stability Selection picked three covariates, with only ``Stage 3'' in common with BVSNLP.  As in the previous dataset, we tested BVSNLP for different choices of $\tau$ in the interval $[0.01,0.9]$, and the same model was selected for all values within this range. The total runtime of BVSNLP for this dataset was around five minutes using 120 CPUs.

Figure \ref{kirpauc} shows the predictive accuracy for the proposed method based on a five-fold cross-validation. The outcomes for GLMNET, ISIS-SCAD and Stability Selection are not displayed in the plot because those methods did not converge or failed to produce results for at least one of the five folds in the cross-validation experiment.  The small AUC values in this plot for $t<1$ warrant comment.  Because there were few events soon after entry of tissue samples into the TCGA database, the AUC for early timepoints falls close to the 50\% benchmark reflecting no predictive value. %According to \citet{blanche}, a misspecified model could misleadingly result in higher values for other measures of predictive accuracy, such as Harrell's c-index. This is not the case for the time dependent AUC, which is assumption free. {\color{green} I don't understand the previous two sentences.  Could you expand a bit?} This makes AUC a preferred choice for predictive accuracy measurement, but potentially causes AUC to have low values, even close to a 50\% benchmark, {\color{green} for early time points when few deaths are expected.  THIS ADDITION OKAY?}.

\begin{figure}[!t]
	\centering
	\begin{subfigure}{.5\textwidth}
		\centering
		\includegraphics[width=1\linewidth]{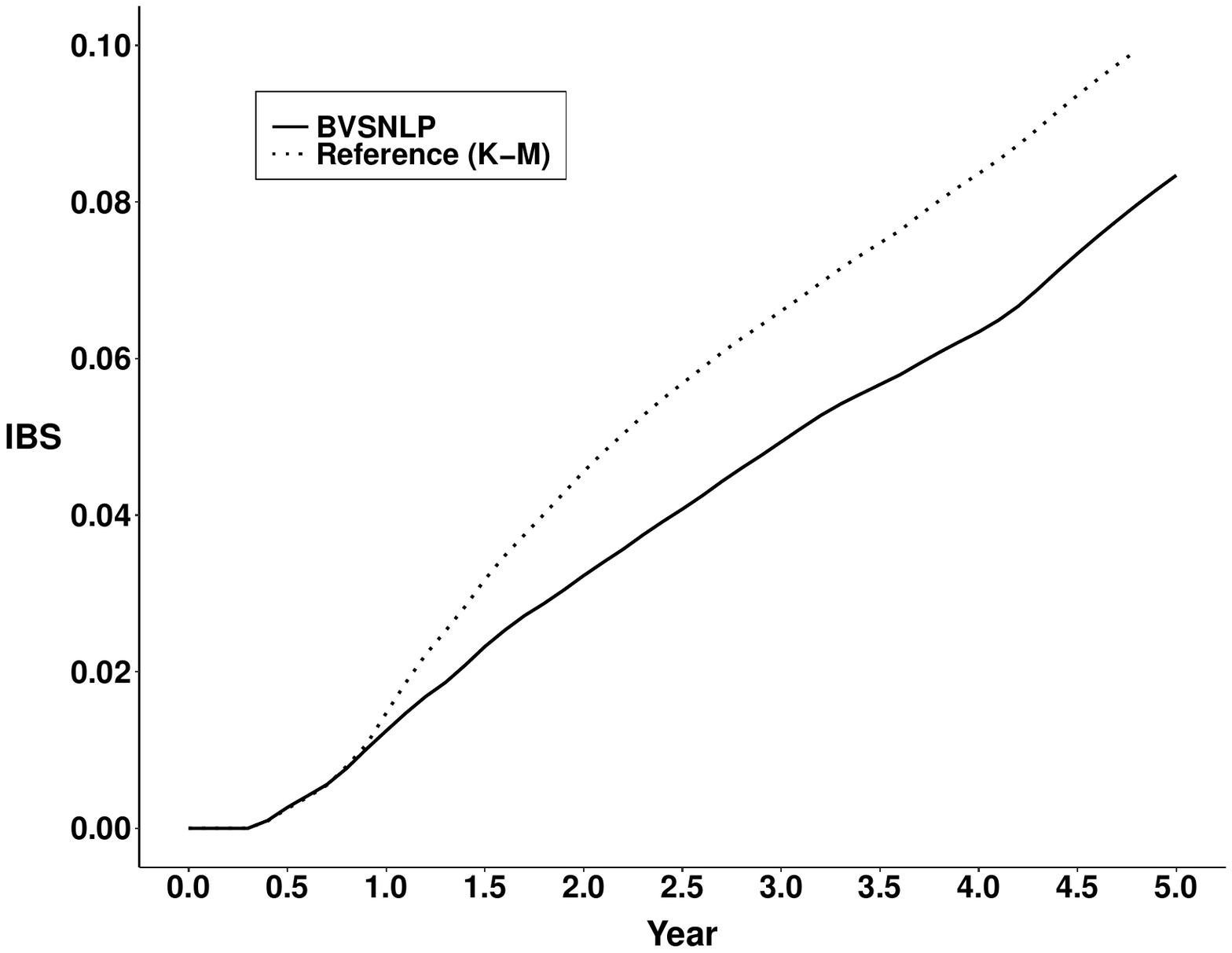}
		\caption{IBS}
		\label{ibskirp}
	\end{subfigure}%
	\begin{subfigure}{.5\textwidth}
		\centering
		\includegraphics[width=1\linewidth]{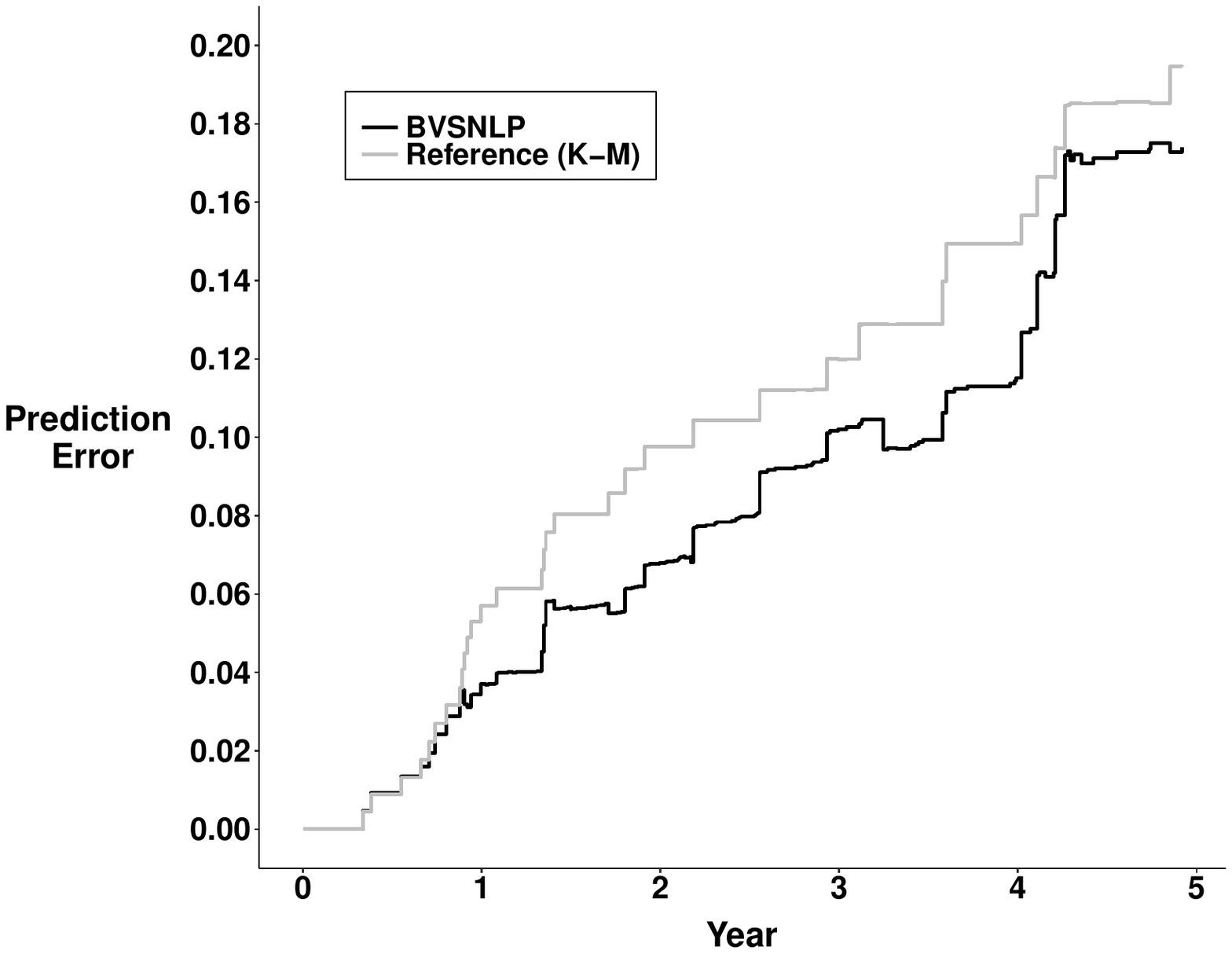}
		\caption{Prediction Error}
		\label{perrkirp}
	\end{subfigure}
	\caption{IBS and prediction error of BVSNLP for the KIRP dataset.}
	\label{perribskirp}
\end{figure}

Figures \ref{ibskirp} and \ref{perrkirp}, respectively, depict IBS and prediction error curves of the BVSNLP method, based on a five-fold cross validation for the KIRP dataset, and compares it to the reference curve obtained by the Kaplan--Meier method. The c-index measure for the BVSNLP method is 0.876. The average runtime for BVSNLP in each fold of the cross-validation was 3.6 minutes on 120 CPUs.

%\begin{figure}[!t]
%		\centering
%		\includegraphics[width=80mm]{ibs_kirp}
%		\caption{IBS of BVSNLP for the KIRP dataset.}\label{ibskirp}
%\end{figure}
%
%\begin{figure}[!hb]
%		\centering
%		\includegraphics[width=80mm]{perr_kirp}
%		\caption{Prediction error of BVSNLP for the KIRP dataset.}\label{perrkirp}
%\end{figure}

%================================================================================================
%================================================================================================

\section{Discussion}\label{c4:discuss}

In this article a Bayesian variable selection method, BVSNLP, was proposed for selecting variables in high and ultrahigh dimensional datasets with survival time as outcomes. BVSNLP uses an inverse moment nonlocal prior density on nonzero regression coefficients. Analyses of simulated and real data suggest that BVSNLP performs comparably or better than other existing methods for variable selection for survival data. Moreover, the real data results indicated that the proposed algorithm is robust to the choice of the hyperparameter $\tau$ in the piMOM prior for values of $\tau$ in the range $[0.01, 0.9]$.

Various outputs are provided by the algorithm. These include the HPPM, MPM and the posterior inclusion probability for each covariate in the model. For real datasets, Bayesian model averaging is used to incorporate uncertainty in selected models when computing time dependent AUC plots using Uno's method \citep{survuno}. Finally, an R package named BVSNLP has been implemented to make the algorithm freely available and adaptable to interested researchers. The package can be run in parallel fashion where hundreds of CPUs can be exploited in order to increase the number of visited models in the search for the highest posterior probability model. The BVSNLP package is available in the R repository, CRAN, at \href{https://CRAN.R-project.org/package=BVSNLP}{https://CRAN.R-project.org/package=BVSNLP}. The user manual for the package is also available from this site.

Two real cancer genomic datasets from the TCGA website were considered in this article. Compared to other methods, BVSNLP found sparser models with biologically relevant genes. The proposed method showed a reliable predictive accuracy as measured by AUC using substantially fewer variables.

We have based our assessments on time dependent AUC and biological interpretation of the results, but other measures, like IBS, prediction error and the concordance index (also known as the c-index or Harrell's c-index) were also reported. Difficulties associated with such measures are identified in \citet{blanche}.  In particular, the authors of that article demonstrate that the concordance index can favor misspecified models over the correctly specified model because it is based on the order of event times rather than the event status at the prediction horizon. This may explain the slightly higher c-index values for GLMNET in both simulation and real datasets. The time dependent AUC does not suffer from this deficiency. Of course, different evaluation criteria can be expected to result in different rankings of models, and criteria that emphasize prediction error over low false positive rates can be expected to favor larger models. Similarly, criteria that place a higher premium on eliminating false positives will tend to select smaller models.

% === For the response letter.
%As alternatives to AUC, other measures of predictive performance such as Integrated Brier Score (IBS) \citep{ibs} can be used as well. The R package \verb|pec| estimates the Brier score and provides prediction error curves for survival data. However, that package uses maximum likelihood estimates of coefficients for its internal calculations and survival curve estimators. However, for our analysis we use the MAP estimate of the coefficients in estimating the survival curve as discussed in Section \ref{estsurv}. In this article, we presented time dependent AUC plots as a measure of predictive performance and computing IBS for BVSNLP by modifying the code of the \verb|pec| package can be left as future work.

%\newpage
\vspace{15pt}
{\large\textbf{Acknowledgments:}} Portions of this research were conducted with the advanced computing resources provided by Texas A\&M High Performance Research Computing (HPRC). 

%\begin{supplement}
%	\sname{Supplementary Material}\label{suppA}
%	\stitle{``Bayesian Variable Selection For Survival Data Using Inverse Moment Priors''.}
%	\slink[doi]{Completed by the typesetter}
%	\sdescription{The supplementary material to this article containing some of the details regarding the methodology, algorithm for implementing that methodology, and some extended discussion on the results.}
%\end{supplement}

\bibliography{ref}

\end{document}

\maketitle

\section{Calculating the Gradient and Hessian of \texorpdfstring{$\lowercase{g}(\bfbk)$}{Negative Log Posterior Function}} \label{gradhess}
Let $l(\bfy;\bfbk) = \log(\pi(\bfy\con\bfbk))$ and $l_\pi(\bfbk) = \log(\pi(\bfbk))$.
For a $n\times p$ matrix $\bfA$, let $\bfA_{(i)}$ denote the $n\times 1$ vector corresponding to the $i^{th}$ column of $\bfA$ and $\bfA_j$ denote the $1\times p$ vector corresponding to the $j^{th}$ row of $\bfA$. Also let $\overline{\bfA}_i =  (\bfA_{i:n, .})^T$, where $\bfA_{i:n, .}$ is the sub-matrix of $\bfA$ from row $i$ to the last row where all columns are included.  This makes the dimension of $\overline{\bfA}_i$ equal to $p \times (n-i+1)$. Similarly, for a vector $\bfalpha$ of size $n$, let $\overline\bfalpha_i$ denote the sub-vector of $\bfalpha$ components $i,i+1,\ldots,n$, a vector of size $(n-i+1)$.

Let $\psi_{\bfk_i} = \sum\limits_{j=i}^ne^{\bfXk_j\bfbk}$ and $\bfpsi_\bfk=(\psi_{\bfk_1},\ldots,\psi_{\bfk_n})^T$. Also let $\bfeta$ denote the $n \times 1$ column vector $\exp\{\bfXk\bfbk\}$. The logarithm of $\pi(\bfy\con\bfbk)$ in (2.2) can then be expressed as
\begin{equation}\label{c4:coxloglik}
l(\bfy;\bfbk)= \bfdel^T\big(\bfXk\bfbk-\log(\bfpsi_\bfk)\big).
\end{equation}
For each $n\times k$ design matrix $\bfXk$ and $\bfbk$ vector, define a new $k \times n$ matrix $\wXk$, with $i^{th}$ column
\begin{equation}\label{c4:xstardef}
\wXk_{(i)}=\frac{(\barbfXk_i)(\overline\bfeta_i)}{\psi_{\bfk_i}}.
\end{equation}
Here, $\barbfXk_i$ and $\overline\bfeta_i$ are obtained from matrix $\bfX_\bfk$ and vector $\bfeta$, respectively, using the notation described in the beginning of this section.

The negative gradient of $l(\bfy;\bfbk)$ can then be written as
\begin{equation}\label{c4:loglikderv}
-\frac{\partial l(\bfy;\bfbk)}{\partial \bfbk} = \big[\wXk-\bfXk^T\big]\bfdel.
\end{equation}

To compute the Hessian matrix, let $\mathbb{X}_{\bfk_{ij}}$ be the $(i,j)$ element of $\wXk$. The $k\times k$ identity matrix is denoted by $\mathbf{I}_k$ and $D(\bfalpha)$ denotes a diagonal matrix with the elements of the vector $\bfalpha$ on its diagonal. Finally, let $\bfzeta^j=\bfXk_{(j)}$ denote the $j^{th}$ column of $\bfXk$.

Row $j$ of the $k \times k$ Hessian matrix of $-l(\bfy;\bfbk)$ is defined as
\begin{equation}\label{c4:hess1}
-\frac{\partial^2 l(\bfbk)}{\partial\bfbk_j\partial\bfbk^T} = \bfdel_{1\times n}^T\bfOmega^j_{n\times k}.
\end{equation}

The $\bfOmega^j_{n\times k}$ matrix itself is constructed row by row, with row $i$ equal to 
\begin{equation}\label{c4:hess2}
\bfOmega^j_i = \bigg[\barbfXk_i\frac{D(\overline\bfzeta^j_i)}{\psi_{\bfk_i}}\overline\bfeta_i - \wXk_{ji}\wXk_{(i)}\bigg]^T.
\end{equation}
Computing the Hessian can be implemented with a computational complexity of $O(n)$.

The gradient and Hessian of the logarithm of the piMOM prior is more straightforward, and is given by
\begin{equation}\label{c4:priorgrad}
-\frac{\partial l_\pi(\bfbk)}{\partial \bfbk_i} = \frac{r+1}{\bfbk_i} - \frac{2\tau}{\bfbk_i^3}.
\end{equation}
The Hessian of $-l_\pi(\bfbk)$ is a diagonal matrix, $D(\bfalpha)$, where
\begin{equation}\label{c4:priorhess}
\alpha_i = \frac{6\tau}{\bfbk_i^4}-\frac{r+1}{\bfbk_i^2}.
\end{equation}

%======

\bigskip

\section{Estimating Individual Survival Curves}\label{estsurv}
In the Cox proportional hazard model the survival function for individual $i$ under model $\mathcal{M}_\bfk$ is defined as
\begin{equation}\label{survfunc}
S_i(t;\mathcal{M}_\bfk) = \exp{\big\{-H_0(t;\mathcal{M}_\bfk)\big\}}^{\exp\{\bfxki^T\bfbk\}}
\end{equation}
where $H_0(t;\mathcal{M}_\bfk) = \int_0^t h_0(\tau;\mathcal{M}_\bfk)d\tau$ is the cumulative baseline hazard function, which can be estimated by

\begin{equation}\label{breslow}
\widehat{H}_0(t;\mathcal{M}_\bfk) = \sum_{i:\; t_i \leq t}\frac{\delta_i}{\sum\limits_{j=i}^n \exp\big\{\bfxkj^T\bfbk\big\}}.
\end{equation}
This is known as the Breslow estimator of $H_0(t;\mathcal{M}_\bfk)$ (the observed times are sorted as in (2.2)). At this point three approaches can be exploited to estimate the survival curve for individual $i$. The first approach is to compute the HPPM survival curve by replacing $\mathcal{M}_\bfk$ with $\mathcal{M}_\text{{HPPM}}$, and use the MAP estimate of $\bfb$ under the HPPM, $\hat{\bfb}_\text{{HPPM}}$, in (\ref{survfunc}) and (\ref{breslow}). That is,

\begin{equation}\label{survest1}
\widehat{S}_i(t) = \exp{\big\{-\widehat{H}_0(t;\mathcal{M}_\text{{HPPM}})\big\}}^{\exp{{\{\bf X}_{\text{HPPM}_i}^T\widehat{\bfb}_\text{HPPM}\}}}.
\end{equation}

The second approach is computationally more intensive but takes into account the uncertainty of the posterior samples of the model space. In this approach, samples from the posterior distribution of the survival function are generated by replacing $\bfk$ in (\ref{survfunc}) with every posterior sample of the model space. The estimated survival curve is then obtained by taking the average of the posterior samples. That is,

\begin{equation}\label{survest2}
\widehat{S}_i(t) = \frac{1}{\mathcal{K}}\sum\limits_{j=1}^\mathcal{K} \exp{\big\{-\widehat{H}_0(t;\mathcal{M}_{\bfk_j})\big\}}^{\exp\{{\bf X}_{{\bfk_j}_i}^T\widehat{\bfb}_{\bfk_j}\}},
\end{equation}
where $\mathcal{K}$ is the number of posterior samples.

The third approach is to use Bayesian model averaging. As discussed in the previous section, we use Occam's window where only the models with posterior probability of at least $0.01\times p(\mathcal{M}_\text{{HPPM}}|\bfy_n)$ are used in model averaging. Suppose $\mathcal{N}$ models fall in Occam's window. Then

\begin{equation}\label{survest3}
\widehat{S}_i(t) = \sum\limits_{j=1}^\mathcal{N} \exp{\big\{-\widehat{H}_0(t;\mathcal{M}_{\bfk_j})\big\}}^{\exp\{{\bf X}_{{\bfk_j}_i}^T\widehat{\bfb}_{\bfk_j}\}}p(\mathcal{M}_{\bfk_j}\con \bfy_n).
\end{equation}

%======

\section{Comparison With pMOM Prior}
\begin{figure}[!ht]
	\centering
	\includegraphics[width=80mm]{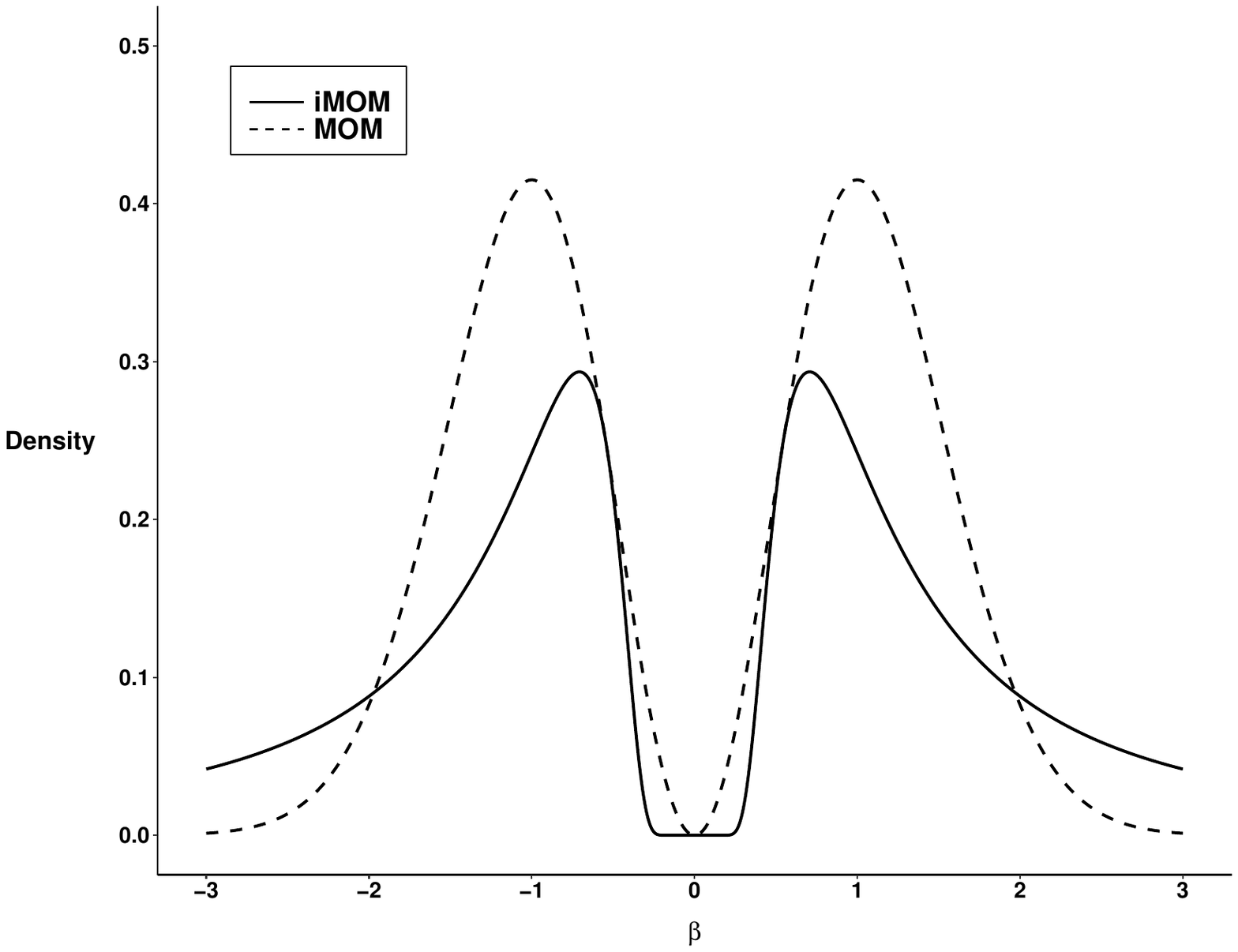}
	\caption{iMOM and MOM prior with $r=1$ and $\tau=0.5$.}\label{pimom}
\end{figure}

Another nonlocal prior that might be considered as a potential candidate for the prior densities on the non-zero coefficients in model $\bfk$ is the product of independent MOM priors, or the pMOM densities \citep{johnsonrossel}, specified by
\begin{equation}\label{pmom}
\pi(\bfbk|\tau,r)=(2\pi)^{-k/2}\tau^{(-rk-k/2)}\exp\Big(-\frac{\mathbf{\bm{\beta}^\prime_k}\bfbk}{2\tau}\Big)\prod_{i=1}^k\beta_i^{2r}, \quad r\in\mathbb{N},\tau>0.
\end{equation}
The hyperparameter $\tau$ has the same role as in piMOM densities in (2.8) and $r$ is the order of the density. An example of a MOM prior for $r=1$ and $\tau=0.5$ is depicted in Figure \ref{pimom}. % This plot illustrates the heavy tail of the iMOM density as well as its behavior around zero in contrast to the MOM density.

Following the discussion of nonlocal priors in Section 2.3, we again note that for $r=1$ and a fixed $\tau$, piMOM densities assign negligible probability to a wider region around zero than do  pMOM densities.  More specifically, pMOM densities decrease to zero at only an inverse polynomial rate while piMOM densities decrease at a rate that is order $\exp(-\tau/\beta^2)$, which is much faster.  Consequently, smaller false positive rates are expected for procedures based on piMOM priors that those based on pMOM priors.  On the other hand, pMOM densities have tails that converge to zero at an exponential rate, while piMOM densities have heavier, Cauchy-like tails. Consistency properties of piMOM priors for linear models were studied in \citet{s5}. In that setting it was shown that piMOM priors are consistent for $p=O(e^{n^\nu}), 0<\nu<1$. By comparison, this property does not hold for pMOM priors. The source of inconsistency for pMOM priors in $p\gg n$ settings stems from the fact that their densities go to zero only at an inverse polynomial rate in a neighborhood of the origin. For these reasons, piMOM-based procedures are more effective for variable selection in $p\gg n$ settings.

To better demonstrate the practical importance of these properties, we performed 20 simulation studies where the number of observations and covariates were $n=200$ and $p=10,000$. The true model had size 6 with true coefficients equal to $(0.5, 0.85, 1.00, 1.50, 1.85, 2.5)$.  The sign of coefficients were chosen randomly with probability $0.5$ in each simulation. The columns of the design matrix were multivariate Gaussian random variables with mean $0$ and marginal variance of $1$. The correlation between every two variables was $0.5$. In each of the simulations, we fixed $r=1$ and assigned $\tau =$ 15 different values in the interval $[0.01, 10]$. The survival times were simulated from an exponential distribution with mean $10$.

The proposed variable selection algorithm was implemented on the simulation data using both pMOM and piMOM priors. Table \ref{tauvals} summarizes the outcome of the selection procedure using different hyperparameter values for both priors. The numbers are averaged over 20 simulations. In that table, MTPR is the mean true positive rate, MFPR is the mean false positive rate and TMP is the proportion of times that the true model was found without any false positives.

\begin{table}[!t]
	\centering
	\caption{Comparison of variable selection outcomes between pMOM and piMOM for different values of hyperparameter, $\tau$, over 20 simulations.}\label{tauvals}		
	\begin{tabular}{c|ccccccccc}
		\hline
		\rule{0pt}{3ex}\multirow{2}{*}{Hyperparameter $\tau$} & & \multicolumn{2}{c}{MTPR(\%)}& & \multicolumn{2}{c}{MFPR(\%)}& & \multicolumn{2}{c}{TMP}\\ \cline{3-4} \cline{6-7} \cline{9-10}
		\rule{0pt}{3ex}& & piMOM & pMOM & & piMOM & pMOM & & piMOM & pMOM \\\hline
		\rule{0pt}{3ex}0.01 && 100 & 22.50 && 0 & 0.15 && 1 & 0\\
		0.2 && 100 & 21.67 && 0 & 0.12 && 0.9 & 0\\
		0.4 && 99.17 & 21.67 && 0.01 & 0.12 && 0.9 & 0\\
		0.6 && 99.17 & 20.83 && 0.01 & 0.12 && 0.9 & 0\\
		0.8 && 97.50 & 20.83 && 0.01 & 0.12 && 0.85 & 0\\
		1.0 && 95 & 20.83 && 0 & 0.12 && 0.70 & 0\\
		1.25 && 95 & 20.83 && 0 & 0.12 && 0.70 & 0\\
		1.6 && 94.16 & 20 && 0 & 0.12 && 0.65 & 0\\
		2.0 && 93.33 & 20 && 0 & 0.12 && 0.60 & 0\\
		2.3 && 93.33 & 19.17 && 0 & 0.12 && 0.60 & 0\\
		3.8 && 90.83 & 18.33 && 0 & 0.12 && 0.45 & 0\\
		5.4 && 87.50 & 18.33 && 0 & 0.12 && 0.35 & 0\\
		6.9 && 85 & 18.33 && 0 & 0.12 && 0.30 & 0\\
		8.4 && 81.67 & 18.33 && 0 & 0.12 && 0.20 & 0\\
		10.0 && 81.67 & 18.33 && 0 & 0.12 && 0.20 & 0\\\hline    
	\end{tabular}
\end{table}

As shown in Table \ref{tauvals}, the pMOM model never finds the true model for any of the $\tau$ values. Moreover, the average true positive rate for pMOM is always 5 times less than that for piMOM, and the average false positive rate for pMOM is higher than it is for piMOM. This suggests variable selection based on piMOM priors in ultrahigh dimensional settings is likely to perform better than variable selection based on pMOM priors.

\section{Discussion on the parameters of the S5 algorithm}\label{s5param}
The goal of the S5 algorithm is to accurately estimate the HPPM, which is determined as the highest posterior probability among those models visited. Thus, the main objective in our algorithm is to increase the number of visited models.

There are two important parameters in the S5 algorithm. The temperature vector for the annealing schedule, and the number of iterations at each temperature. We use 10 equally spaced temperature values decreasing from 3 to 1, where at temperature $t_l$ the posterior probability is raised to the power of $1/t_l$. Values of $t<1$ increase the posterior probability to unreasonably large values, making S5 susceptible to being trapped in local extremes, this reducing the number of visited models. Therefore, 1.0 is the lowest chosen temperature for the annealing schedule. Values higher than 3, on the other hand, were found empirically to not improve the performance of the algorithm because it then visited too high a proportion of models with comparatively low posterior probability. 

The other parameter, the number of iterations at each temperature, can be chosen by the user in the R package. Theoretically, the higher number of iterations, the more models that will be visited.  For the analyses in this paper, the number of iterations was chosen to be 30. This choice was based on a sensitivity analysis performed on simulation data for different numbers of iteration values ranging from 20 to 50, where we investigated the identification of the HPPM and number of visited models.  The details of this experiment follow. 

We defined a simulation batch as 50 different datasets that were generated with the same settings as Case 3 of the simulations discussed in Section 4, but with true model size of 6 and coefficients equal to $\beta_1= -1.5140, \beta_2=1.2799, \beta_3=-1.5307, \beta_4=1.5164, \beta_5=-1.3020, \beta_6=1.5833$, and $\beta_i=0$ for $i>6$. A run of the BVSNLP was run on each dataset to find the simulation truth. For each simulation batch, in addition to the average number of unique visited models, the proportion of times (out of 50) that the algorithm selected the true model, without any false positives or false negatives, was also recorded. The \verb|niter| parameter of the S5 algorithm was varied for each simulation batch, ranging from 10 to 50 in increments of 5.

\begin{figure}[t]
	\centering
	\includegraphics[width=80mm]{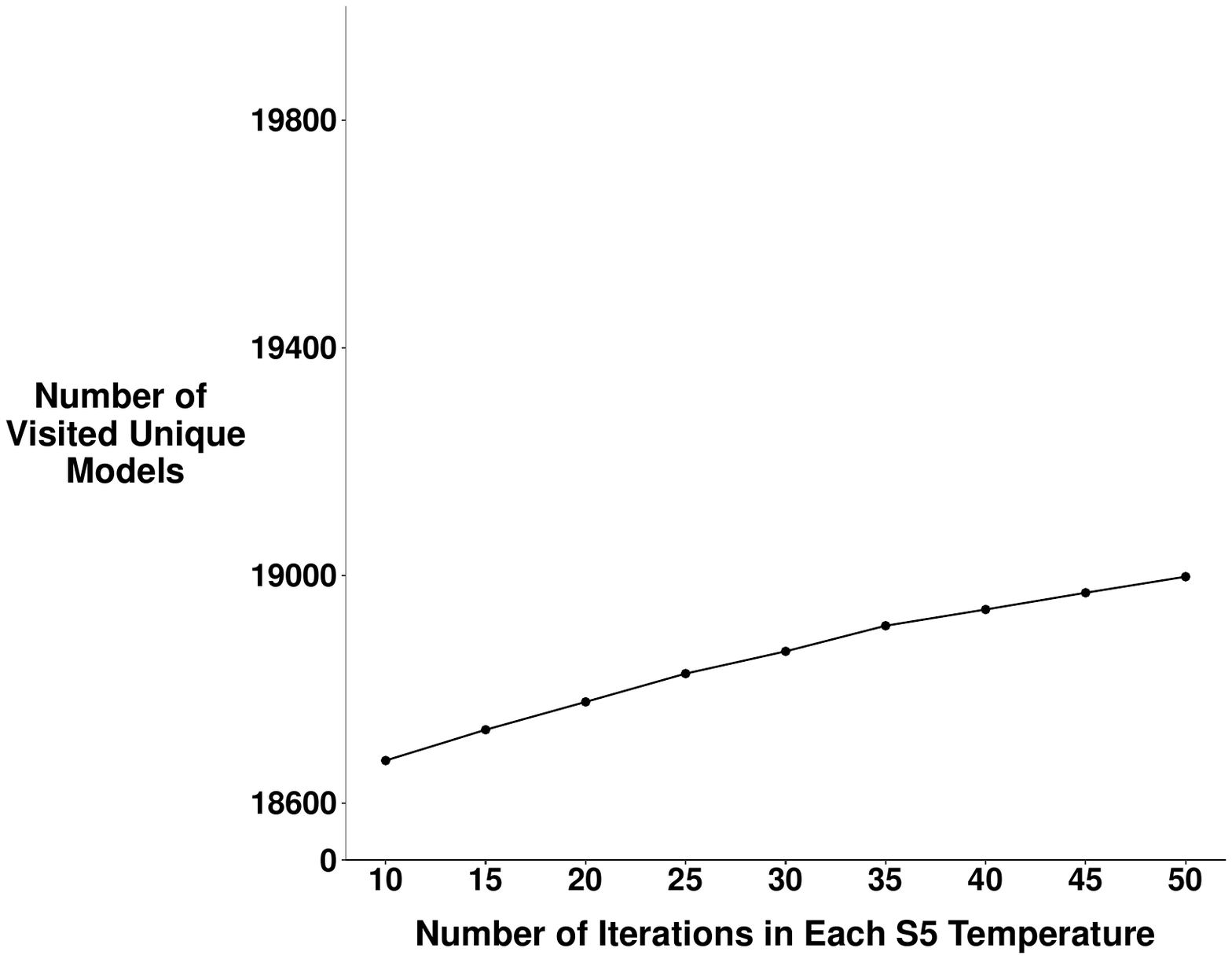}
\caption{Average number of unique visiteds models by BVSNLP for different iterations in S5 algorithm, for simulation case 3.}\label{fig:senvm}
\end{figure}

The outcome of the sensitivity analysis for this parameter of the S5 algorithm is summarized in Figure \ref{fig:senvm} for the average number of unique visited models. The difference between the minimum and maximum average number of visited models in all cases was only 324 models, and the proportion of times the true model was found with no false positives or negatives was 100\% in all simulation cases.

These results suggest the S5 algorithm's performance in finding the true model was not significantly impacted by the parameter \verb|niter|, and that the number of visited unique models just changed $1.73\%$ from an average of 18,674.78 unique models in 10 iterations per temperature to 18,997.64 unique models in 50 iterations per temperature in the S5 algorithm. This experiment suggests that the BVSNLP algorithm is relatively insensitive to the parameters of the S5 stochastic search algorithm, at least within the range of values considered in this study.

%\begin{table}[!h]
%		\centering
%		\caption{Proportion of times the true model is found for different iterations in S5 algorithm, for simulation Case 3.}\label{senT}
%		\begin{tabular}{l|ccccccccc}
%			\hline
%			\rule{0pt}{3ex}\verb|niter| & 10 & 15 & 20 & 25 & 30 & 35 & 40 & 45 & 50 \\\hline
%			\rule{0pt}{3ex}$P$ & 100\% & 100\% & 100\% & 100\% & 100\% & 100\% & 100\% & 100\% & 100\% \\\hline		
%		\end{tabular}
%\end{table}

\section{Definitions of MTP and MFP}\label{mtpdef}
Let $S_i$ be the set of all covariates selected as the final model by the method at iteration $i$. Also let $\bfk$ be the set of covariates in the true model.

Define
\begin{equation}
\begin{split}
&(\text{TP})_i = |S_i\cap\bfk|, \;\text{and }\\
&(\text{FP})_i = |S_i\setminus\bfk|, %\;\mbox{and }\\
% &(P)_i = I(TP_i == |\bfk|\; \&\& \; FP_i==0),
\end{split}
\end{equation}
where $|A|$ denotes cardinality of set $A$ and $A\setminus B$ denotes the set minus operation.%, and $I(.)$ is an indicator function which is 1 if the condition inside the parentheses is satisfied and 0 otherwise. In the argument of the indicator function above, $"=="$ is the equality operator and $"\&\&"$ is the logical $and$ operator. 

Following the definitions above, MTP and  MFP are obtained as follows: %and P are obtained as follows:
\begin{equation}
\text{MTP} = \frac{1}{m} \sum\limits_{i=1}^m (\text{TP})_i\;; \;\; \text{MFP} = \frac{1}{m} \sum\limits_{i=1}^m (\text{FP})_i, %\;; \;\; \mbox{P} = \frac{1}{m} \sum\limits_{i=1}^m (P)_i,
\end{equation} 
where $m$ is the total number of iterations.

\newpage
\section{Expression Regulation Networks for The KIRC Dataset}
Figure \ref{genepairs} depicts the regulation networks relative to the analysis of the KIRC dataset in Section 5.1 of the main manuscript.  This illustration shows that the genes identified by the survival analysis and logistic regression analyses have strong functional relationships.

\begin{figure}[!ht]
	\centering
	\subfloat[]{{\includegraphics[width=90mm]{genepair1} }}%
	\qquad
	\subfloat[]{{\includegraphics[width=90mm]{genepair2} }}%
	\caption{Expression regulation networks connecting the old and new gene sets. a) This diagram shows all genes that are in the shortest pathways through the expression regulation between SAV1 and AR. b) This diagram shows all genes in the shortest pathways through expression regulation between NUMBL and SUDS3.}%
	\label{genepairs}%
\end{figure}

%\bibliography{ref}